\documentclass[revtex4-2]{aastex62}

\usepackage{savesym}
\usepackage{amsmath}
\usepackage{booktabs}
\usepackage{natbib}
\usepackage{comment}
\usepackage[colorinlistoftodos]{todonotes}

\submitjournal{AJ}

\usepackage{bm}
\usepackage{cancel}
\usepackage{enumerate}
\usepackage{amssymb}
\usepackage{mathtools}
\usepackage{comment} 

\usepackage{subcaption}
\savesymbol{tablenum}
\usepackage{siunitx}
\DeclareSIUnit{\sol}{
 \text{$\text{M}_{\odot}$}
}

\restoresymbol{SIX}{tablenum}

\shorttitle{Modeling Color-Magnitude Diagrams with Neural Flows}
\shortauthors{Cranmer et al.}

\renewcommand{\tt}[1]{\texttt{#1}}
\newcommand\photg{\texttt{phot\_g\_mean\_mag}}

\usepackage{cleveref}
\begin{document}

\title{Modeling the Gaia Color-Magnitude Diagram with Bayesian Neural Flows\\
to Constrain Distance Estimates}

\author{Miles D. Cranmer}
\affiliation{Department of Astrophysical Sciences, Princeton University, Princeton, NJ 08544, USA}
\author{Richard Galvez}
\affiliation{Center for Data Science, New York University, 60 Fifth Avenue, New York, NY 10011, USA}
\affiliation{Center for Cosmology and Particle Physics, Department of Physics, New York University, 726 Broadway, New York, NY 10003, USA}
\author{Lauren Anderson}
\affiliation{Center for Computational Astrophysics, Flatiron Institute, 162 Fifth Avenue, New York, NY 10010, USA}
\author{David N. Spergel}
\affiliation{Center for Computational Astrophysics, Flatiron Institute, 162 Fifth Avenue, New York, NY 10010, USA}
\affiliation{Department of Astrophysical Sciences, Princeton University, Princeton, NJ 08544, USA}
\author{Shirley Ho}
\affiliation{Center for Computational Astrophysics, Flatiron Institute, 162 Fifth Avenue, New York, NY 10010, USA}

\correspondingauthor{Miles D. Cranmer}
\email{mcranmer@princeton.edu}

\begin{abstract}
We demonstrate an algorithm for learning a flexible
color-magnitude diagram from noisy parallax
and photometry measurements
using a normalizing flow, a deep
neural network capable of learning
an arbitrary multi-dimensional probability distribution.
We present a catalog of 640M photometric distance posteriors to
nearby stars derived from this data-driven model
using Gaia DR2 photometry and parallaxes.
Dust estimation and dereddening is done iteratively inside the model and without prior distance information, using the Bayestar map.
The signal-to-noise (precision)
of distance measurements improves on average by more than
48\% over the raw Gaia data, and we also
demonstrate how the accuracy of distances have improved over
other models, especially in the noisy-parallax regime.
Applications are discussed, including significantly improved Milky Way
disk separation and substructure detection. We conclude with a discussion of future work, which
exploits the normalizing flow architecture to allow us to exactly marginalize
over missing photometry, enabling the inclusion of many surveys without losing coverage.
\end{abstract}


\section{Introduction} \label{sec:intro}

Gaia's precise astrometry
has impacted the astrophysics community
in countless ways.
While Gaia data is predominantly useful for mapping out the Milky Way
and its bulk properties, for example in \cite{bovy_stellar_2017};
it has been used for calibration of standard
candles like the Red Clump as in \cite{hawkins_red_2017,huber_asteroseismology_2017},
and vice versa --- using the Red Clump to improve
Gaia parallax calibration --- \cite{hall_testing_2019}; and
mapping out the interstellar medium to create dust maps
of the Milky Way, as in \cite{green_3d_2019}.
Gaia's astrometric data has been incredibly useful
for the study of substructures in the Milky Way
and its halo, for example, stellar streams,
as in \cite{price-whelan_off_2018,malhan_ghostly_2018,brown_gaia_2018,koposov_piercing_2019}.
The precise positional and proper motion estimates
even allow for 
automated algorithms
to be used to detect stellar streams,
as described in \cite{malhan_streamfinder_2018}.
Recently the astrometric data has been used
to observe possible dynamical evidence for dark matter in stellar
streams in \cite{bonaca_spur_2019}, although this stream
was at a distance where the parallax measurements couldn't
be reliably used besides a foreground filter.
Producing more accurate astrometry for the Gaia DR2 dataset
would impact many current and future use cases.
Gaia parallaxes greatly degrade in quality
at a distance starting at about $\sim\SI{1}{kpc}$, with only
$\sim 72$ million sources having greater than 10 signal-to-noise (SNR, defined here as
parallax over parallax uncertainty),
hampering studies of structure in the galactic halo such as stellar
streams. 

We can make reasonable distance estimates from these uncertain
parallaxes by relying on distance priors such as those
derived in \cite{bailer-jones_estimating_2015,bailer-jones_estimating_2018},
or by exploiting patterns in stellar photometry, either using
theoretical Color-Magnitude Diagrams (CMDs) for stars or learning them as in \cite{leistedt_hierarchical_2017,anderson_improving_2018}.
If we were to rely on purely theoretical models for
estimating distances from photometry, we would be subject
to systematic errors from the models themselves.
Distance prior-derived distances are also heavily prior-dominated,
and of limited use for substructure studies, also demonstrated
in Section~\ref{sec:applications}.
By denoising in a model-independent fashion ---
training a flexible machine learning algorithm to 
model the most common photometric measurements for stars, hence
acting as a prior for distance estimates ---
we avoid difficulties from modeling the theoretical photometry,
and naturally learn over top of the Gaia systematic
errors. Using a very flexible model for a color-magnitude
population density would reduce potential model-dependent bias
from entering distance estimates.

One way of building very flexible models is by using machine learning models.
Typically in astronomy,
machine learning is thought of purely as an approach to classification
or regression,
which is predicting an output scalar or vector value
based on a set of input parameters, such as the simplest example:
creating a line of best fit.
The extension of this is
to fitting quadratic functions, and then higher-order polynomials,
and other simple models,
which can be generalized to multi-input, multi-output using
additional parameters.
One generalization of these simple machine learning
models which are fit to data
is a popular technique called deep learning, which
is a flexible approach capable of fitting very complex
surfaces over input data, relying typically on stochastic gradient
descent to fit millions of parameters. Deep learning
can also be applied to many additional optimization problems than just regression,
such as density estimation, which we will use it for in this paper.
Density estimation is the problem of fitting a function that models
an unknown probability distribution and can be approached
with deep learning using a model called a normalizing flow.

Deep learning can be thought of as a
recursive generalized linear regression --- you
repeatedly compute linear regression (fitting a hyperplane)
on an input, following each regression
with an element-wise nonlinearity (such as converting negative values to zero).
In the case of normalizing flows, as we will see,
we also apply a mask over the linear regression weights to make the function
have a triangular Jacobian matrix.
We use such a normalizing
flow in this paper to model photometric measurements of Gaia stars.

One common criticism of deep learning models targets their
lack of interpretability and potential over-flexibility.
Generally, it is recommended that deep learning should only be used
when it gives you large performance gains
or lets you model a relation
that would be extremely difficult to represent
with classical machine learning models. As discussed in this paper,
due to the large size of the Gaia DR2 dataset, and the non-Gaussian contours
of the density of stars on a CMD,
a deep neural network is a useful model for its scalability and flexibility,
so we choose it over traditional machine learning models.

\section{Data}\label{sec:data_description}

Our dataset is a slice of the Gaia DR2 catalog.
For technical papers describing the 2nd data release used in this paper,
consult \cite{brown_gaia_2018,lindegren_gaia_2018,riello_gaia_2018}.
We make use of the following features:
\texttt{bp\_rp}, \texttt{bp\_g}, \texttt{phot\_g\_mean\_mag}, \texttt{ra},
\texttt{dec}, \texttt{parallax},
along with their corresponding uncertainties,
and features to calculate the renormalized unit weight error (RUWE):
\texttt{astrometric\_chi2\_al} and
\texttt{astrometric\_n\_good\_obs\_al}.
We choose to not apply any filters based on parallax or parallax uncertainty,
meaning we take all noisy measurements (although we do filter
using RUWE, discussed later), including negative parallaxes.
We exclude all stars below -30 deg declination,
since the Bayestar dust map, which we use from
\cite{green_galactic_2018} in the package \cite{dustmaps}, does not extend there.
We also exclude stars lying in holes
of this dust map.
We include all stars during training, even the low-SNR
parallaxes. However, we make cuts during training such
that stars in high-density regions of the sky are excluded
to avoid stars with bad goodness-of-fit for parallax. We do this by requiring
that the renormalized unit-weight error of measured parallaxes,
discussed on \url{https://www.cosmos.esa.int/web/gaia/dr2-known-issues#AstrometryConsiderations},
is less than 1.4.

We apply a single global parallax offset of
0.029 mas from
\cite{lindegren_gaia_2018} to all of the Gaia data.
While the true parallax offset is conditional on several variables,
we believe that training with a constant parallax offset
makes it easier to apply different post-processing offsets
to the finished model. A more complex
parallax offset can also be used during evaluation.

\section{Model}\label{sec:model}

\newcommand{\parmod}{\varpi_\text{model}}
\newcommand{\parobs}{\varpi_{\text{obs}}}
\newcommand{\partrue}{\varpi_{\text{true}}}
\newcommand{\erpar}{\sigma_\varpi}

We wish to train a model that takes
\texttt{bp\_rp}, \texttt{bp\_g}, \texttt{phot\_g\_mean\_mag}, \texttt{ra},
\texttt{dec}, \texttt{parallax} from the Gaia DR2 catalog,
along with their uncertainties,
and produces a Bayesian posterior over distance.
Our model is similar in strategy, but not in terms of the actual density model,
to \cite{anderson_improving_2018}. \cite{anderson_improving_2018}
built a model on Gaia DR1 parallaxes cross-matched with 2MASS photometry, 
and applies the ``Extreme Deconvolution'' algorithm from \cite{bovy_jo_extreme_2011}
to build a Gaussian Mixture Model (GMM) over color-magnitude
values. This GMM is then used as a prior for the same data, to build a final parallax
probability distribution for every data point. The model described in 
this paper is similar to this GMM model but differs
by using a normalizing flow to model density in photometric space.
Various other changes are made, such
as that we propagate uncertainty from the dust map, both into
our training (down-weighting photometry from uncertain dust estimates)
and during evaluation (sampling the dust map).

Normalizing flows are not yet popular for density estimation
in astrophysics or the natural sciences in general, although
they are being used for some likelihood-free inference applications derived
from \cite{papamakarios_sequential_2018}
in Cosmology and Particle Physics, 
for example in the ``MadMiner'' package for LHC data \cite{brehmer_madminer:_2019},
and ``PyDELFI'' for Cosmology \cite{alsing_fast_2019}.
This paper represents the first application that we are aware of
for normalizing flows being applied to learning a stellar CMD.

Our model makes several assumptions
which largely follow those made in \cite{anderson_improving_2018}:
\begin{itemize}
    \item The fundamental assumption of the model is that for a given star,
        there will be other stars with similar photometry, as is assumed
        for any regression-type optimization problem.
    \item We model the contents of Gaia DR2, not the Milky Way.
        Since parallax measurements are noisier for stars in the halo, which also tend to be lower metallicity,
        the CMD model will bias to metallicities in the disk. However, we lessen
        the effect of
        this by training on all stars, rather than just high-SNR.
    \item The model assumes that the expected color-magnitude relation
        is unchanging with respect to location in the Milky Way,
        in that we do not include $\alpha, \delta$
        (right ascension and declination) as a parameter in our CMD.
    \item This model makes no further physical assumptions about stellar structure
        or the color-magnitude relation. This is a trade-off because
        while it does avoid potential inaccuracies of physical models, it also
        misses the many successes of these models. Future work will attempt
        to combine physical models with data-driven to help regularize the training.
    \item We assume that the Bayestar dust map of \cite{green_galactic_2018} produces
        accurate dust posteriors along each line of sight. We further
        assume that for training, the median dust estimate
        will not bias the model during training. However, we incorporate the full
        dust posterior for evaluation. 

    \item We make the assumption that the Gaia catalog has 
        no parallax bias relative to any parameters
        (such as proper motion, or a specific sky coordinate such as 
        a high-density region).
\end{itemize}
A potentially problematic assumption in our model is the fact that the dust map we use, from \cite{green_galactic_2018},
makes use of stellar models to estimate dust, meaning
that there are implicitly some stellar models being introduced in our distance
estimates. In the future, it would be
desirable to combine the learning of a dust map with the learning of a CMD
so we could make completely model-independent estimates,
but for now, this is our approach.
The assumptions for the model described in our paper are different from those in \cite{anderson_improving_2018}
in that we do not model the color-magnitude relation using
128 Gaussians; rather, we use a deep normalizing flow which has more
flexibility in describing the joint prior.
The path of stellar evolution has many sharp turns and discontinuities, so,
while the spread of stars about a single isochrone may be Gaussian-like 
due to a Gaussian spread in ages and metallicities, the overall CMD
is better modelled with a highly flexible model that can express sharp turns,
which is where deep learning can be extremely helpful,
as GMMs perform poorly at modeling the contours of the CMD.
We also deal with a significantly larger dataset, using the majority of DR2 (640M stars) versus
a selection of DR1 (1.4M stars).
We also incorporate the entirety of
the samples from \cite{green_galactic_2018} rather than median dust estimates
during evaluation.

\begin{figure*}
\centering
\includegraphics[width=0.6\textwidth]{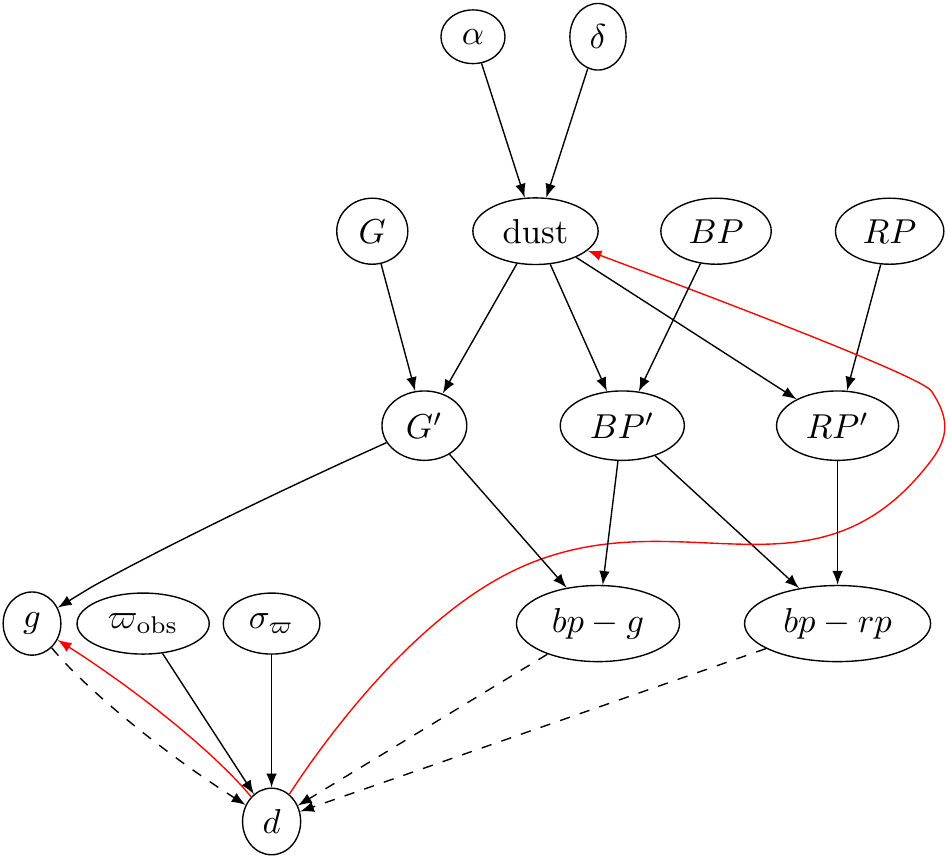}
\caption{A Bayesian graphical model
for distance estimates in our model. Gaia data is used
in combination with a normalizing flow model of the color-magnitude
diagram to generate a Bayesian posterior over distance.
The dashed lines, which represent
the three-dimensional color-magnitude diagram, are those
learned by the normalizing flow described in this paper.
The red paths are the update step: the current
expectation value for distance ($d$)
is used to update the absolute $g$-magnitude
for each star as well as the dust value integrated
along the line of sight. This iteration occurs 5 times during
training, and 10 times during evaluation. An optional prior can be applied
during training and evaluation that is conditional on 
$\alpha, \delta$.}
\label{fig:graphicalmodel}
\end{figure*}

The crux of our model is a very flexible ``normalizing flow'' neural
network. A good introduction on this family of model can be found at \url{http://akosiorek.github.io/ml/2018/04/03/norm_flows.html}.
This architecture models the joint posterior density for Gaia magnitudes,
as
\[ P(g, bp-rp, bp-g), \]
where $g$ is the absolute (dereddened) G-band mean magnitude from Gaia DR2, corresponding
to column \photg,
and $bp$ and $rp$ similarly for the absolute integrated BP and RP mean magnitudes.
This prior density models the dashed lines in the graphical model in
\cref{fig:graphicalmodel}.

This posterior models the density of stars in Gaia DR2 and
weights stars observed in Gaia DR2 by
\[ \frac{1}{\sigma_P^2}, \]
where we define $\sigma_P$ as
\[ \sigma_P^2 = \sigma_g^2 + \sigma_{bp-rp}^2 + \sigma_{bp-g}^2, \]
which is a metric for the signal-to-noise ratio of data points,
much like when calculating a mean from uncertain measurements,
one weights each measurement by the inverse variance.
These $\sigma_i^2$ values are the variance of estimates in each of the
quantities. These variances incorporate both
the uncertainty of the dust map as well as the
intrinsic uncertainty in the Gaia parallax and distance
prior.

During training, we sample 32 parallaxes for each Gaia DR2 data point
from the truncated normal distribution:
\begin{equation}
\label{eqn:distance}
P(\varpi) \propto \left\{
\begin{array}{cc}
    \exp\left(-\frac{(\varpi-\varpi_\text{obs})^2}{2 \sigma^2}\right),
     & \varpi>0 \\
 0,     & \varpi\leq0,
\end{array}
\right.
\end{equation}
using the DR2
parallax value for $\varpi_\text{obs}$, and $\sigma$ equal
to the DR2 standard deviation in the parallax value estimate.
As stated in \cite{hogg_likelihood_2018}, a likelihood for Gaia parallaxes
treats the individual measurements as drawn from $\mathcal{N}(\varpi_\text{true}, \sigma)$.
Therefore this truncated distribution acts as a minimal distance prior
on the Gaia parallax likelihood, which states that we can only have positive distances.

\subsection{Model architecture}
\label{sec:arch}

A normalizing flow can roughly be thought of
as a Gaussian that has been parametrically warped
by an invertible neural network. This
is a smooth invertible mapping between probability distributions:
$\mathbb{R}^n \rightarrow \mathbb{R}^n$.
Our specific model relies on the ``Masked Autoregressive Flow'' variant,
described in \cite{papamakarios_masked_2017} for density
estimators, which uses the following mappings:
\begin{align*}
    y_1 &= \mu_1 + \sigma_1 z_1,\\
    \forall i>1: y_i &= \mu(y_{1:i-1}) + \sigma(y_{1:i-1}) z_i,
\end{align*}
for $y, z, \mu \in \mathbb{R}^n$ and $\sigma\in\mathbb{R}^n_{+}$.
This equation is similar to
the common matrix multiply followed by vector addition that is found
in neural networks, but with a particular mask applied. This is done
so that the Jacobian of this mapping is triangular --- and hence
the determinant is easy to calculate as the product of elements along
the diagonal.
Recall that if you would like to change variables from $\vec{x}$ to $\vec{y}$
via a smooth function $f$, for probability distributions one must write:
\[ p(y) = p(x) \abs{\frac{df}{dx}}^{-1} \]
This mapping is invertible, and the inverse also has a triangular Jacobian.
The fact that the Jacobian is easy to calculate lets us normalize the 
transformation between probability distributions.
The inverse is
\begin{align*}
    \forall i>1: z_i &= \frac{y_i - \mu (y_{1:i-1})}{\sigma(y_{1:i-1})},
\end{align*}
which transforms from our data variables ($y$) to a latent
space ($z$) where we set the Gaussian. This transform is 
what we compute to calculate the probability
of given data (though we use a reparametrized version,
so it doesn't need to be done sequentially).
This transform can be repeated to form a complex flow.
Autoregressive models can also be used to exactly
marginalize over inputs in the case of missing data (such as missing photometry),
as discussed in Section~\ref{sec:future}.

Our model uses a sequence of blocks of 
\textsc{MADE} $\rightarrow$ \textsc{BatchNorm} $\rightarrow$ \textsc{Reverse},
where \textsc{MADE} is the 
Masked Autoencoder for Distribution Estimation
model defined in \cite{germain_made:_2015},
\textsc{Reverse} is the reversing layer found in \cite{dinh_density_2016},
and \textsc{BatchNorm} is a batch norm-like layer (typically
used in convolutional neural networks) also defined for normalizing flow
models in \cite{dinh_density_2016}. We use the PyTorch code
found at \url{https://github.com/ikostrikov/pytorch-flows/blob/master/flows.py}
as the template of our codebase, which we alter
for our usecase.

The \textsc{MADE} model essentially applies
three densely-connected neural network layers with
a mask (hence, a \underline{m}asked \underline{a}utoencoder)
applied to the weights at each layer to satisfy
properties of the neural flow.
It forms an autoencoder such that
each of the output units only depends on the
preceding input units ($i \sim 1:i-1$).
One can think
of this transform as parametrizing an arbitrary
bijective vector field, where the vectors show
the flow of points from distribution to distribution.
The \textsc{BatchNorm} is the equivalent of a batch normalization
for normalizing flows, and \textsc{Reverse} permutes
the order of the probability variables since
the \textsc{MADE}'s mask treats each slightly differently (e.g.,
$y_i$ depends on $z_{1:i-1}$, whereas $y_1$ only on $z_1$). Hence
the \textsc{BatchNorm} and \textsc{Reverse} layers help regularize training.
Defining one block as
\textsc{MADE} $\rightarrow$ \textsc{BatchNorm} $\rightarrow$ \textsc{Reverse},
our model has a probability distribution with $3$ variables. We
then specify a hidden dimension size in each \textsc{MADE} of
$500$ and use $35$ sequential blocks.

\subsection{Dust estimation}
\newcommand{\ra}{\texttt{ra}}
\newcommand{\dec}{\texttt{dec}}

We build the Bayestar dust map from \cite{green_galactic_2018}
into the model with an iterative approach
using the software package \texttt{dustmap} from \cite{dustmaps}.
We use extinction values based on a \SI{7000}{K} blackbody integrated
over an $R_V=3.1$ dust model from \cite{odonnell_rnu-dependent_1994}, using
the extinction package from \cite{barbary_extinction_2016},
which gives us conversions from the Bayestar dust map to Gaia DR2 bands:
\begin{align*}
    A_G &= 2.71 E(g-r)\\
    E(BP-RP) &= 0.85 E(g-r)\\
    E(BP-G) &= 0.39 E(g-r).
\end{align*}

In the future, we would like to expand this graphical model
with an estimate for the temperature and other stellar
parameters of each star, and incorporate a map of $R_V$.
We would also like to fit a separate graphical model which
learns the best extinction to maximize likelihood over
the data.

Training and evaluation take different
approaches due to the computational expense.
During training,
we calculate a distance by dividing each of the
32 parallax samples from \cref{eqn:distance}. We 
calculate the mean of these distance samples to get
the current best-estimate for a distance.
For every best-estimate distance, we query
Bayestar at the center \ra\ and \dec\ position.
We convert this reddening into each of the Gaia
bands, 
giving us estimates for $G, BP,$ and $RP,$
which gives us $bp-rp$ and $bp-g$.
Then, using each of the 32 samples for parallax,
we convert the $G$ estimate into 32 samples for
$g$.
Next, using the current model for $P(g, bp-rp, bp-g)$,
we calculate the probability of each
$(g, bp-rp, bp-g)$ tuple.
These likelihood values are treated as weights,
and the weights are used to calculate
a new best-estimate for distance via a weighted
sum:
\begin{equation} \label{eqn:weight_sum}
d_\text{best} = 
\frac{\sum_i d_i P(g_i, bp-rp, bp-g)}
{\sum_i P(g_i, bp-rp, bp-g)},
\end{equation}
where each of the $d_i$ is a distance sample.
This $d_\text{best}$ is then fed back into
the loop and a new reddening is found using Bayestar.

This iteration is repeated 5 times during training,
due to the computational expense, but 10 times during evaluation.
Once the final $d_\text{best}$ is given, we calculate
a best-estimate value for the dust. We use this to get
final estimates for the dereddened $G, bp-rp, bp-g$. We then calculate
$g$ using the raw parallax samples and dereddened $G$.
Note that we do not use $d_\text{best}$ to calculate
the final $g$ since this could create a feedback
loop for the probability density and create
unphysical artifacts, which we
experimentally observed.
A final $g_\text{best}$ is then found by averaging
the $g_i$.
The standard deviation of the $g_i$
is used to calculate $\sigma_g$:
\[
\sigma_g^2 = \text{Var}(g_1, \ldots, g_{32}) + \sigma_{G, \text{Bayestar}}^2,
\]
where $\sigma_{G, \text{Bayestar}}$ is the uncertainty in the dust
map for $G$ band at the given distance and sky location. We can
add these variances because $G$ and $g$ are linearly
related.
We could also choose to estimate $g_\text{best}$ by multiplying
each of the weights in
\cref{eqn:weight_sum} by $P(g, bp-rp, bp-g)$, though we found
using it created artifacts in the density which did not go
away with further training.

Using $d_\text{best}$ and
the dust map we calculate the dereddened:
$(bp-rp)_\text{best}$
and
$(bp-g)_\text{best}$.
We also calculate the uncertainty due to these colors:
\begin{align*}
\sigma_{bp-rp}^2 &= \sigma_{bp-rp, \text{Bayestar}}^2,\\
\sigma_{bp-g}^2 &= \sigma_{bp-g, \text{Bayestar}}^2.
\end{align*}
Finally, we are left with
$(bp-rp)_\text{best}, (bp-g)_\text{best},$ and
$g_\text{best}$, along with a measure of the combined uncertainty
of the color-magnitude point $\sigma_P$.
Our loss function for this point is then:
$$-\frac{1}{\sigma_P^2} \log\left\{P\left(g, bp-rp, bp-g\right)_\text{best}\right\}.$$
We sum this over all stars in the DR2 catalog,
after calculating the best-estimate color-magnitude points,
and minimize.

This algorithm learns a very flexible prior
on dereddened $(g, bp-rp, bp-g)$ tuples for stars in Gaia DR2,
which can then be combined with a distance prior and
raw parallax measurements to generate a Bayesian distance
estimate for every star in Gaia. This process is done iteratively
until the dust estimate converges.

\subsection{Model Optimization}

We conduct a hyperparameter search for the normalizing
flow over 80 different models, finding the best
model using Bayesian optimization with summed-log-likelihood
as an optimization metric. The model
is trained on the entirety of Gaia DR2 above \SI{-30}{\degree},
completing several passes over the data with a mini-batch size of 2048.
We found this mini-batch size balances accuracy:
much smaller batch sizes led to the creation of artifacts
in the density map, 
and much larger batch sizes resulted in
early convergence to less accurate models.

During model selection, we randomly initialize the model weights
using Xavier initialization (see \citealt{glorot_understanding_2010})
with a normal distribution, and
train for one pass over Gaia DR2, recording the likelihood
over each random one million-star subset. We explore
the (learning rate, layers, hidden units)
space such that the model fits in a 16 GB GPU, and record
the likelihood as a function of these variables and fractional epochs.
We then pass these measurements to a Gaussian Process with a
radial basis function kernel. We select the next model architecture
by maximizing likelihood plus the uncertainty in the likelihood.
In total, we explore 80 different architectures
and find that 
the best loss was for a model with 35 layers of 500 hidden units each.
We also tried models that used mixtures
of normalizing flows but found these underperformed.
We also found that using the distance prior from \cite{bailer-jones_estimating_2018}
during the estimate of $d_\text{best}$
gave the noisy measurements too much weight,
$\frac{1}{\sigma_P^2},$ since for very noisy measurements,
as $\sigma_\varpi\rightarrow \infty$, the distance prior
and $P(x)$ will be equal, 
but $\sigma_P$ will be finite,
so very noisy measurements will have a large effect on the CMD.

\section{Results}\label{sec:results}

We first demonstrate the ability of this model
to estimate the true color-magnitude diagram
of noisy Gaia data on simulated data
in Section~\ref{sec:simulation}.
Next, in Section~\ref{sec:products} onwards,
we present the results on real data.
The trained normalizing flow is visualized in \cref{fig:cmd},
which shows a clear main sequence and giant branch, with
some color in the log plot indicating it has
learned weight at the white dwarf part of the CMD as well,
unprecedented in previous attempts using full CMD GMM models.
We will release the catalog,
which is described in Table~\ref{tbl:data_summary},
and code from links on
\url{https://github.com/MilesCranmer/public_CMD_normalizing_flow}.

\subsection{Simulation}
\label{sec:simulation}

\begin{figure*}
    \centering
    \begin{subfigure}[b]{0.375\textwidth}
        \centering
        \includegraphics[width=\textwidth]{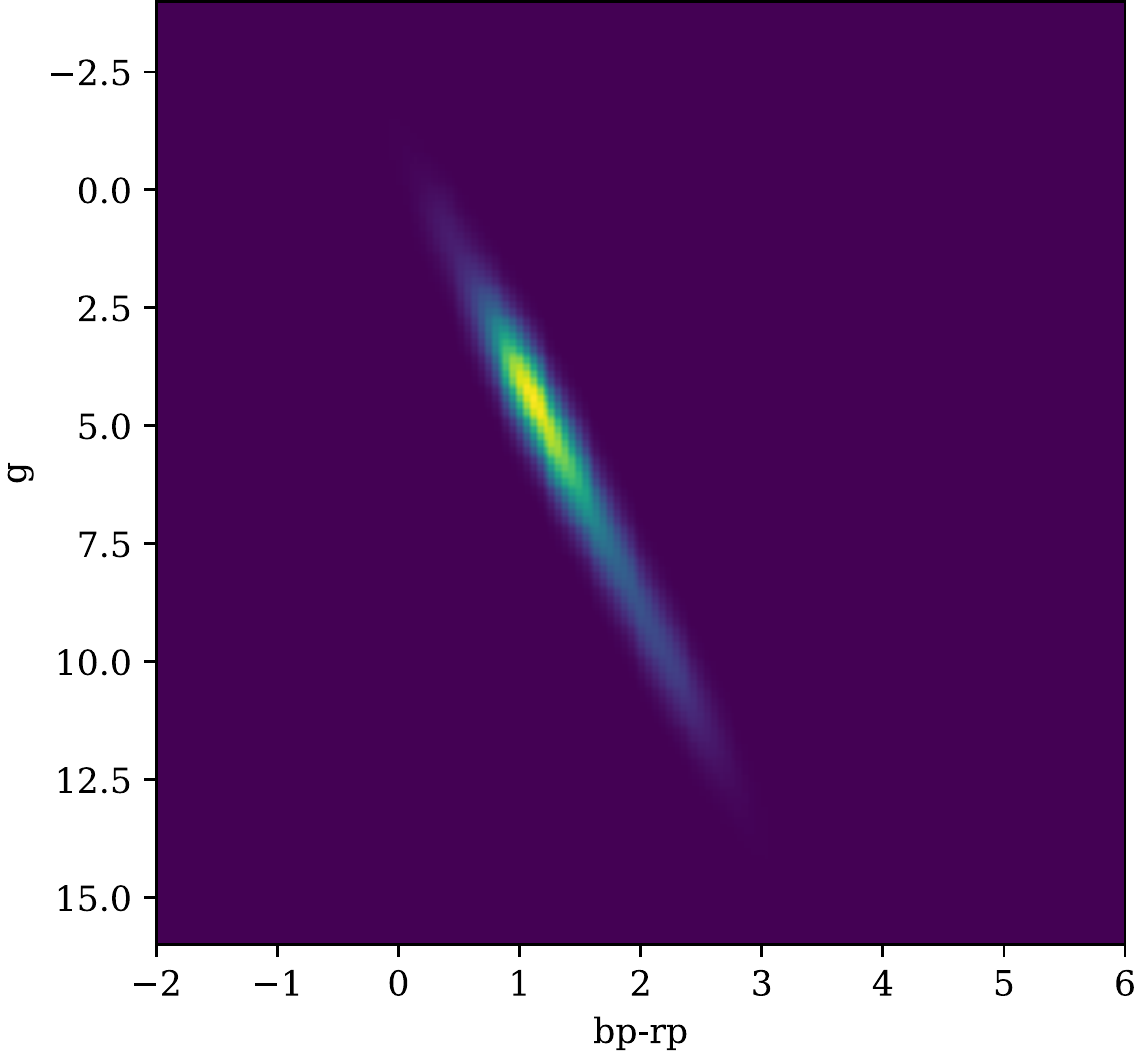}
        \caption[]
        {{\small The true CMD.}}
        \label{fig:true_fake_cmd}
    \end{subfigure}
    \begin{subfigure}[b]{0.375\textwidth}
        \centering 
        \includegraphics[width=\textwidth]{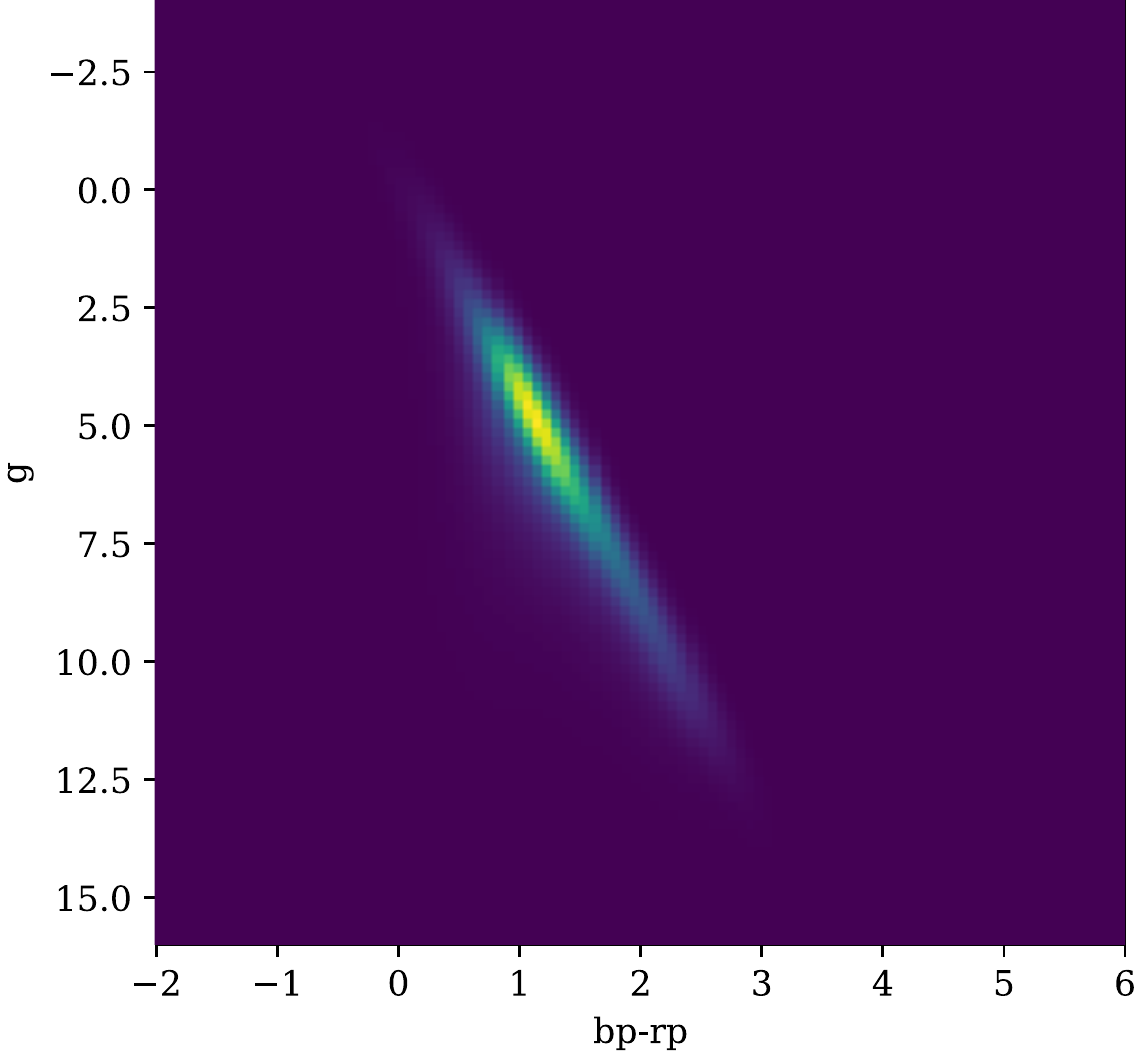}
        \caption[]
        {{\small The reconstructed CMD with our model.}}    
        \label{fig:recon_fake_cmd}
    \end{subfigure}
    \vskip\baselineskip
    \begin{subfigure}[b]{0.375\textwidth}
        \centering 
        \includegraphics[width=\textwidth]{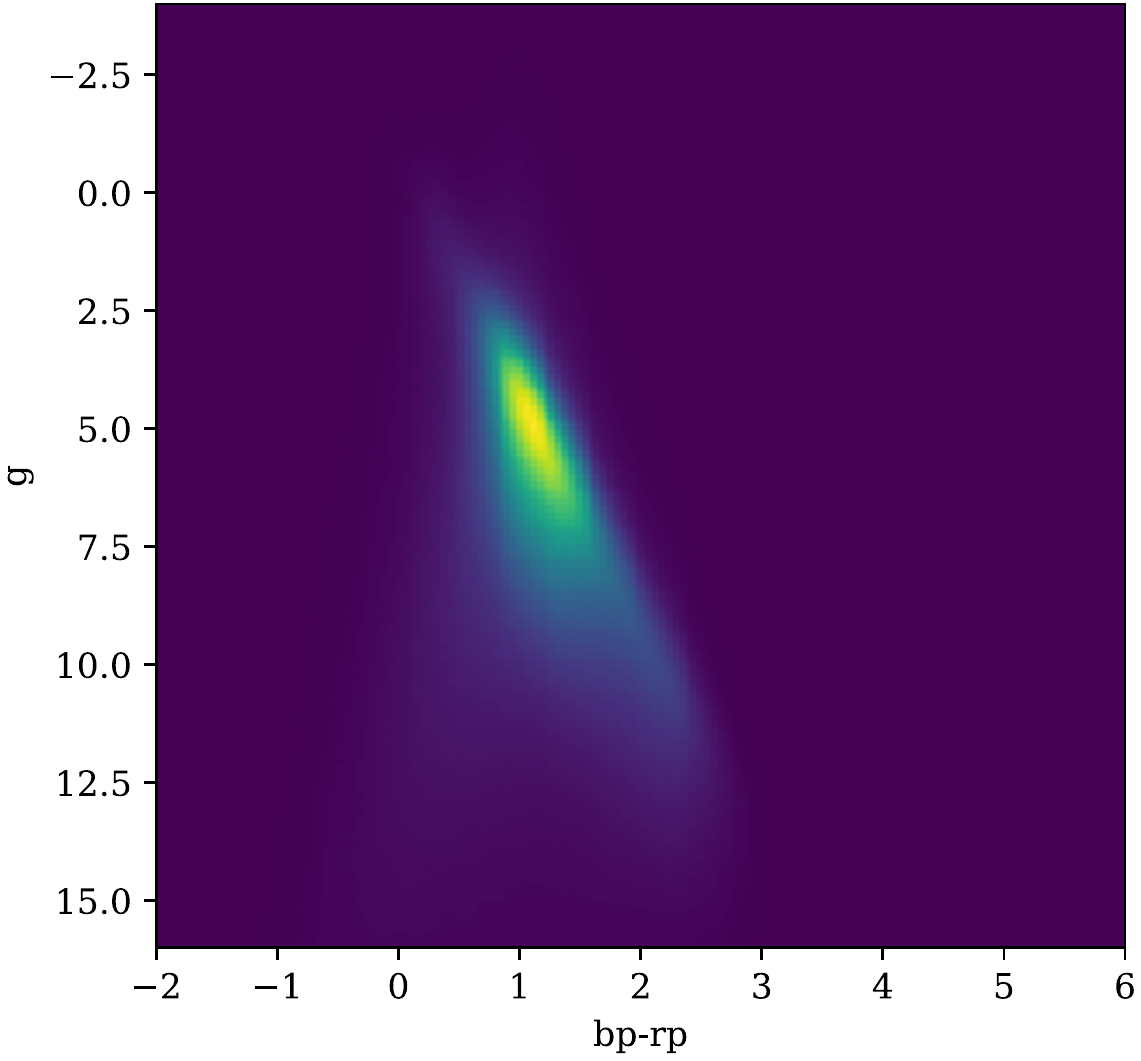}
        \caption[]
        {{\small The observed noisy CMD under dust and noisy parallax estimates.}}    
        \label{fig:noisy_fake_cmd}
    \end{subfigure}
    \quad
    \begin{subfigure}[b]{0.375\textwidth}
        \centering 
        \includegraphics[width=0.9\textwidth]{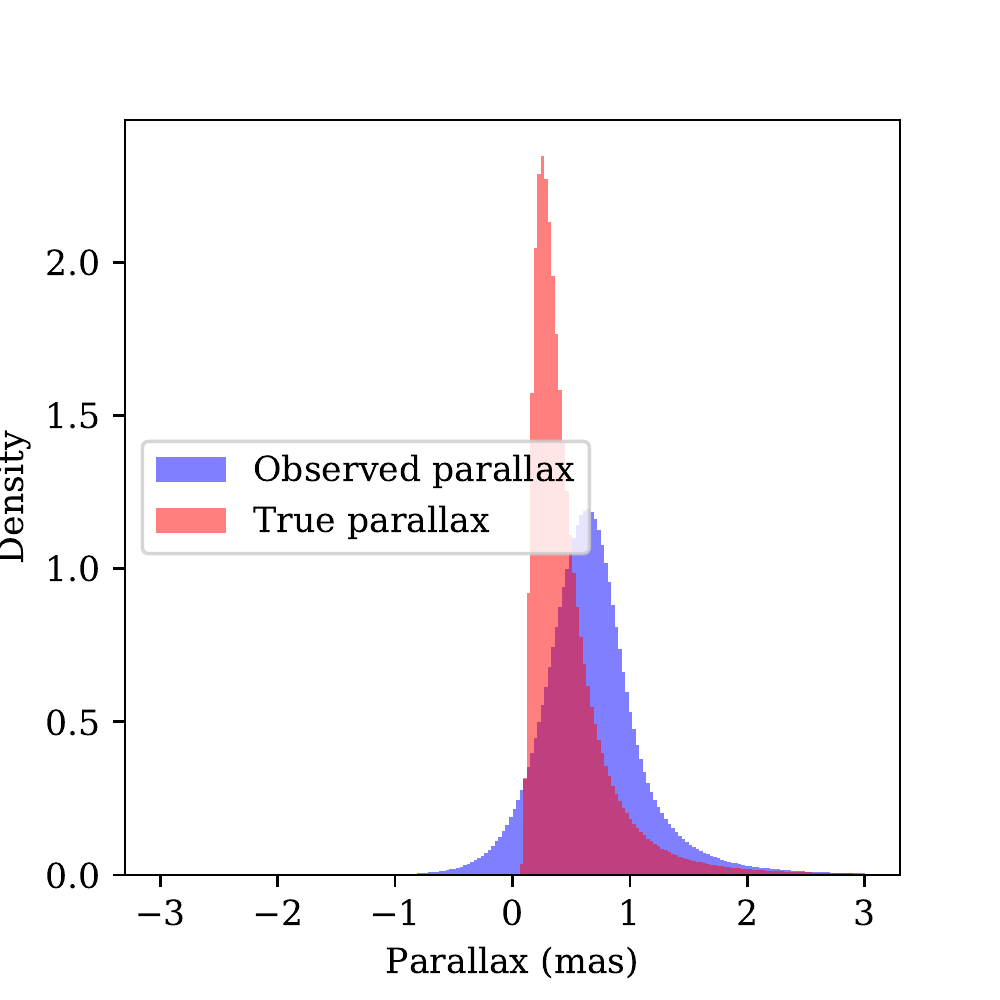}
        \caption[]
        {{\small The distribution of true and observed parallaxes in our simulation.}}    
        \label{fig:fake_cmd_parallax}
    \end{subfigure}
    \caption[]
    {\small A test of our normalizing flow reconstruction of the CMD,
    using a simulated Gaia dataset and a simple fake CMD in plot (a).
    The reconstructed version of this is shown in (b),
    after heavy reddening and noise in the parallax measurements (c).
    }
    \label{fig:fake_dataset}
\end{figure*}

We simulate a basic Gaia dataset
of 30M stars drawn from simple analytic
distributions in distance and color space.
We apply a known noise-free dust map that
is a 2D Gaussian over $(\alpha, \delta)$ that is
uniform along each line of sight.
We distribute the stars uniformly
over each line of sight, $(\alpha, \delta)$ (though the distances
follow a specific prior, see below).
We fit a line in $(g, bp-rp, bp-g)$ space
on the high-SNR Gaia data and use this to
sample colors. We randomly sample
$g$ values from the high-SNR Gaia data
and project onto the line to get
a $bp-rp$ and $bp-g$ value. We then randomly
perturb these colors using a Gaussian
to create the truth CMD shown
in \cref{fig:true_fake_cmd}.
The distances are randomly sampled from the
Bailer-Jones distance prior with $L=\SI{1}{kpc}$:
\[P(d) \propto d^2 \exp{-d/L}. \]
These are used to create the $G$, $BP$, and $RP$
observed bands at Earth,
followed by reddening according to
$d \times \text{dust}(\alpha, \delta)$, for our
dust map, and mapped to fixed extinction conversions
for each band. Next, we use the
formula:
\[ \sigma_\varpi = \frac{d}{\SI{10}{kpc}} \si{mas} \]
to create a parallax measurement error
for all the stars. We then sample
the observed parallax, $\varpi_\text{obs}$, for all stars
from a Gaussian distributed with $\sigma_\varpi$
as the standard deviation. This transformation results
in the adjusted parallax distribution shown
in \cref{fig:fake_cmd_parallax}.
Using $\frac{1}{\varpi_\text{obs}}$ as a simple distance
estimate, we can visualize the noisy reddened CMD
in \cref{fig:noisy_fake_cmd}.

We train a model on this with $8$ blocks
of $256$ hidden nodes with the same
block scheme (MADE, BatchNorm, Reverse)
as our real Gaia model for a few epochs,
allowing the model to use the true dust map
as part of its iterative dust estimation scheme (though
recall it still needs to estimate accurate distances
to calculate the true dust).
We then numerically integrate this CMD
over the $bp-g$ dimension
and visualize it in the same space as \cref{fig:true_fake_cmd}
in \cref{fig:recon_fake_cmd}. As can be seen, the
reconstructed CMD, without any hyperparameter tuning
or extensive training, and noisy reddened data, is very close
to the original.

\subsection{Data Products}
\label{sec:products}

After training the model on the Gaia dataset,
we apply it back to the data to generate a catalog.
We choose not to use the Bailer-Jones distance
priors, since we found they harmed the accuracy
of stellar distance estimates in the halo as they
are very prior dominated (which
can be seen in \cref{fig:gd1_example_prior}).
Instead, we
use a constant distance prior over non-negative distances.
The catalog contains the mean and standard deviation
in the Bayesian posterior estimate
for each of the DR2 sources. We plan on
adding another catalog
that includes 100 quantiles of the posteriors. Alternatively,
one can run our code, which
will be added to \url{https://github.com/MilesCranmer/public_CMD_normalizing_flow}, to sample from the entire posteriors in our trained model.
These will be described in the data
documentation.

\begin{figure}[h]
    \centering
    \begin{subfigure}[b]{0.475\textwidth}   
    \centering
    \includegraphics[width=\textwidth]{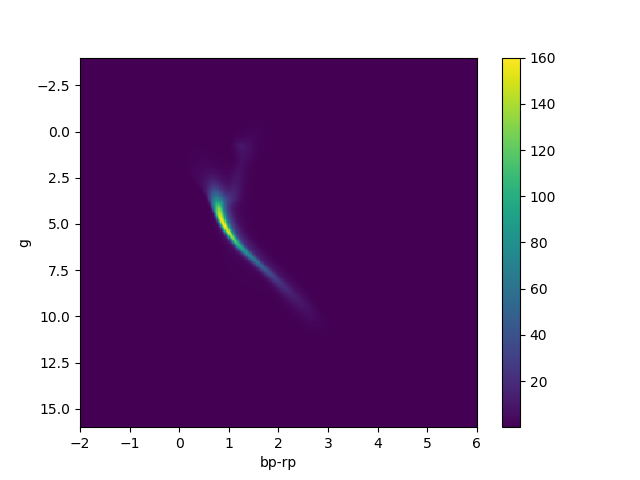}
    \caption{}
    \end{subfigure}
    \begin{subfigure}[b]{0.475\textwidth}  
    \centering
    \includegraphics[width=\textwidth]{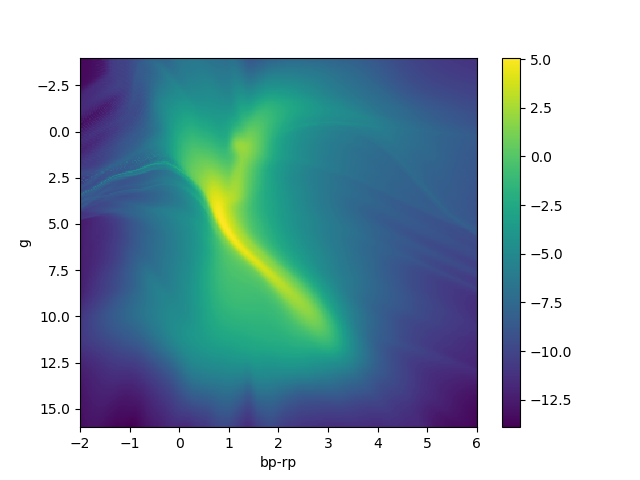}
    \caption{}
    \end{subfigure}
    \caption{The trained normalized flow, representing a color-magnitude diagram,
    is visualized here as a probability density in
    the space of $bp-rp$ and $g$,
    marginalized over $bp-g$. 
    Plot (a) shows probability and plot (b) shows log-probability.}
    \label{fig:cmd}
\end{figure}

\begin{figure*}[h]
    \centering
    \begin{subfigure}[b]{0.8123\textwidth}
        \centering
        \includegraphics[width=\textwidth]{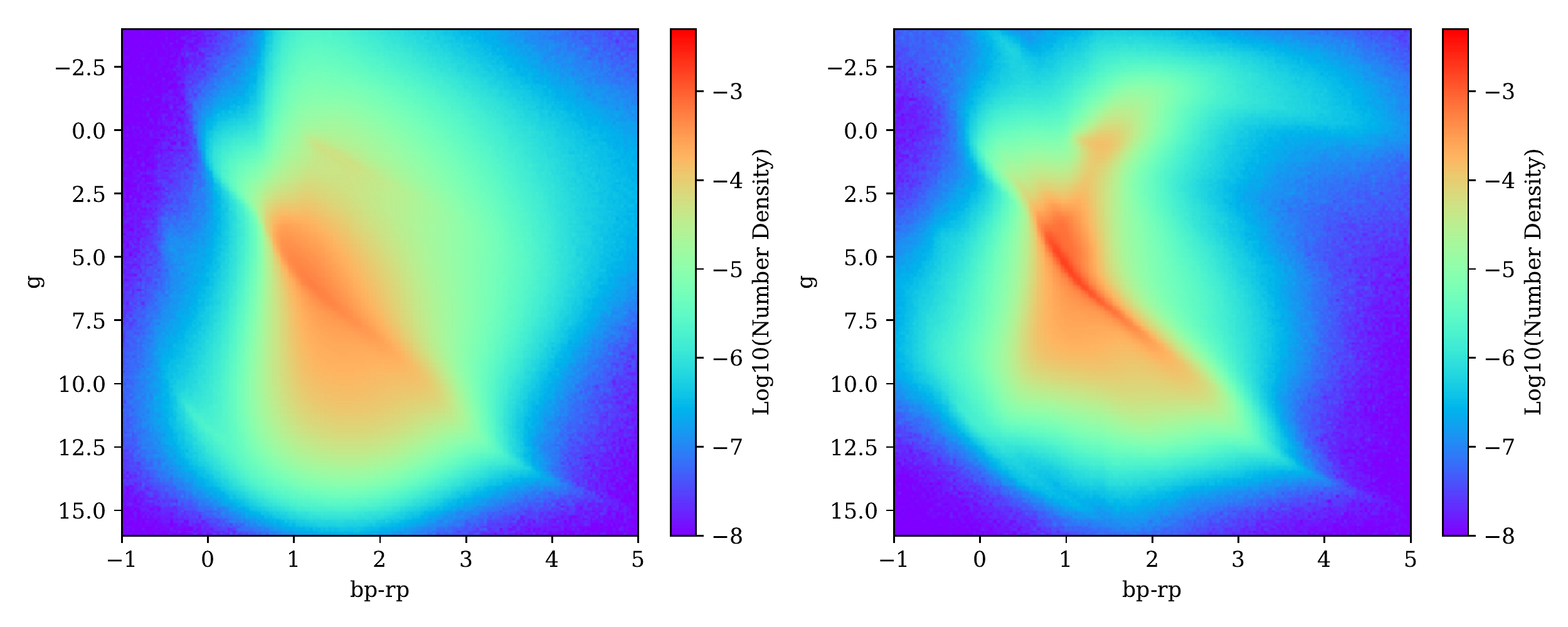}
        \caption{}
        \label{fig:parallaxes}
    \end{subfigure}
    \begin{subfigure}[b]{0.8123\textwidth}
        \includegraphics[width=\textwidth]{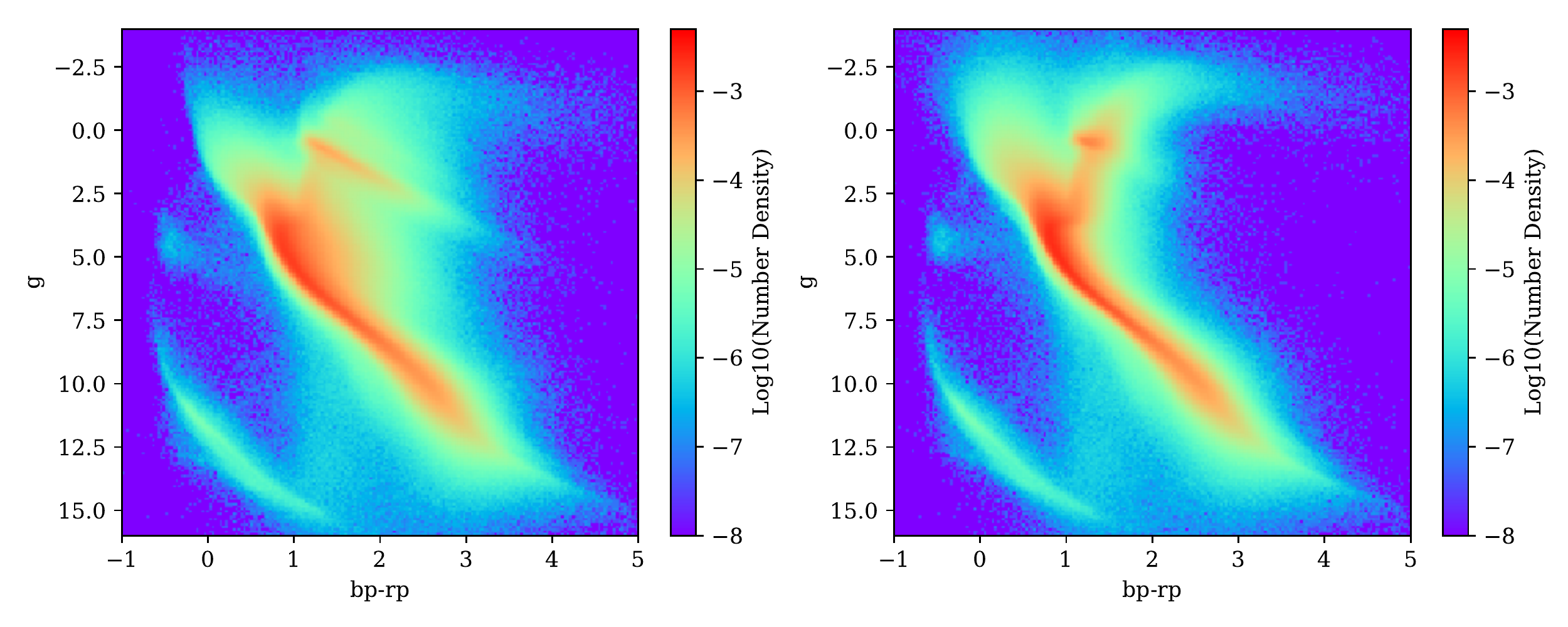}
        \caption{} 
        \label{fig:10_parallaxes}
    \end{subfigure}
    \begin{subfigure}[b]{0.8123\textwidth}
        \includegraphics[width=\textwidth]{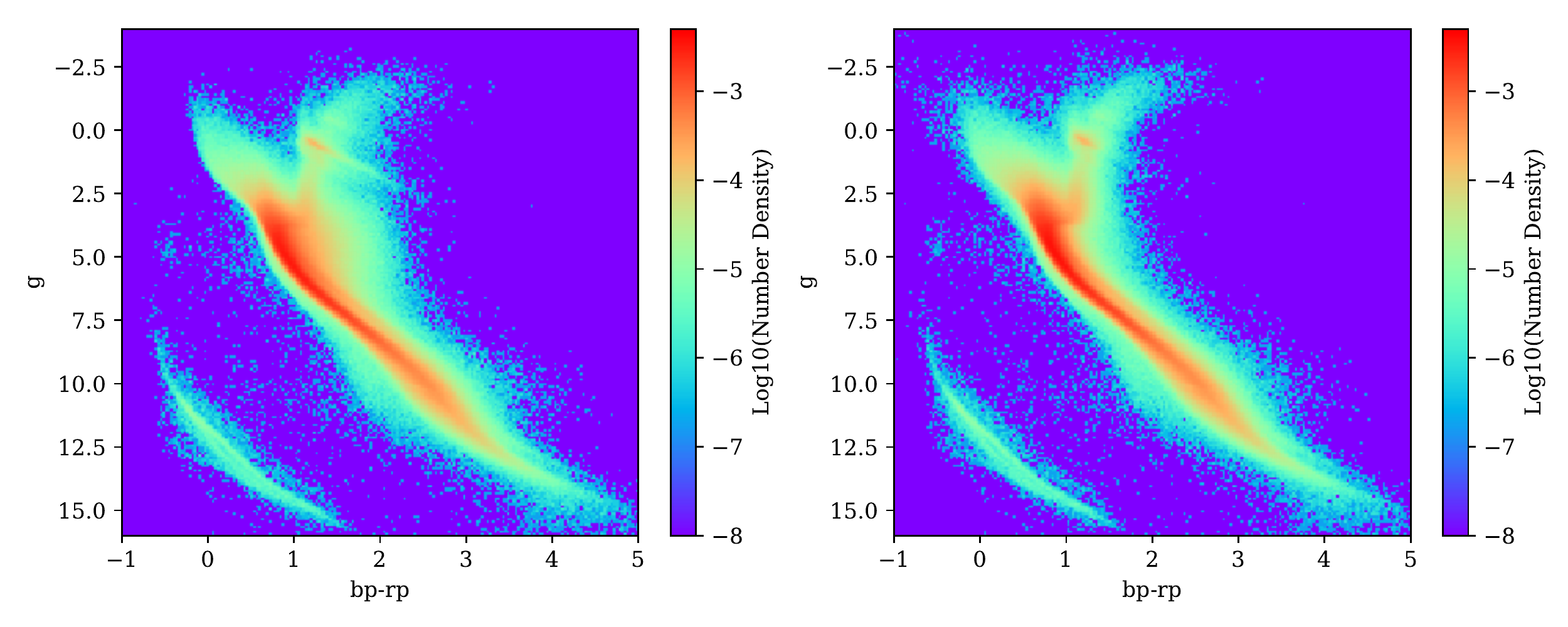}
        \caption{}
        \label{fig:50_parallaxes}
    \end{subfigure}
    \caption{Comparison of Gaia (left) and model (right) color-magnitude diagrams of all stars,
    after denoising the distance estimates using our color-magnitude diagram as a prior.
    The left plots use inverse parallax as distance to calculate $g$, and
    the right plots demonstrate
    the tightening of the CMD around the main and giant branches with
    our model. The logarithm of the normalized number density is represented by color.
    Stars with negative parallaxes 
    are excluded in the left plot, but included in the right plot, which makes the tightening
    of the CMD more impressive. For both, only Gaia stars with
    with RUWE$<1.4$ and dec $>\SI{-30}{\degree}$ are included.
    The plots in (a) have no filter on the stars plotted,
    the plots in (b) have a filter of $\parobs/\sigma_\varpi > 10$ applied,
    and the plots in (c) have a filter of $\parobs/\sigma_\varpi > 50$ applied.
    }
    \label{all}
\end{figure*}

\begin{table*}[h]
    \centering
    \begin{tabular}{@{}lcc@{}}
    \toprule
    Number & Gaia Value (if relevant) & Model Value \\\midrule\midrule
    Total Catalog Entries & &  640,875,169  \\
    \midrule
    Fraction of Stars with $d \in (1 \text{ pc},10 \text{ pc}]$ & & \SI{5.430e-07}{} \\
    Fraction of Stars with $d \in (10 \text{ pc},100 \text{ pc}]$ & & \SI{3.338e-04}{} \\
    Fraction of Stars with $d \in (100 \text{ pc},1 \text{ kpc}]$ & & \SI{1.147e-01}{} \\
    Fraction of Stars with $d \in (1 \text{ kpc},10 \text{ kpc}]$ & & \SI{8.483e-01}{} \\
    Fraction of Stars with $d \in (10 \text{ kpc},100 \text{ kpc}]$ & & \SI{3.659e-02}{} \\
    Fraction of Stars with $d \in (100 \text{ kpc},1000 \text{ kpc}]$ & & \SI{4.261e-06}{} \\
    \midrule
    SNR $\in (-\infty, 0.1]$ & \SI{2.447e-01}{} & \SI{1.650e-04}{} \\
    SNR $\in (0.1, 1.0]$ & \SI{2.162e-01}{} & \SI{1.890e-01}{} \\
    SNR $\in (1 , 10]$ & \SI{4.691e-01}{}  & \SI{7.320e-01}{} \\
    SNR $\in (10 , 100]$ & \SI{6.772e-02}{} & \SI{7.648e-02}{} \\
    SNR $\in (100 , 1000]$ & \SI{2.299e-03}{} & \SI{2.343e-03}{} \\
    Mean SNR Improvement of the points with $\parobs>0$ & &  48.6\%  \\
    \midrule
    Negative Fraction of Parallaxes & \SI{2.229e-01}{} & 0 \\
    Mean SNR of the points with $\parobs < 0$ & &  1.388 \\
    Average Distance of the points with $\parobs < 0$  & & 5.376 kpc \\
    Average Distance (weighted by SNR) of the points with $\parobs < 0$  & & 5.845 kpc \\
    \bottomrule
    \end{tabular}
    \caption{SNR is defined using distance: $d/\sigma_d$, for the model described
    in this paper, and parallax: $\parobs/\sigma_\varpi$, for Gaia. This choice is made because distance
    is the fundamental astrometric measurement of this model, and parallax is the fundamental astrometric measurement
    of the Gaia catalog, and both results are
    given as the parameters of a Gaussian.}
    \label{tbl:data_summary}
\end{table*}
A visualization of the normalizing flow marginalized over $bp-g$
can be seen in terms of probability and log-probability density
in \cref{fig:cmd}.
The tightening of the CMD from applying this prior to the data
can be seen in \cref{fig:parallaxes}
through \cref{fig:50_parallaxes}. The tightening
is used as
a visual metric for the improvement in the
distance estimates.
As seen in Table~\ref{tbl:data_summary}, 
more stars have higher signal-to-noise ratios, 
with only 18.9\% of stars having SNR less
than 1.0 versus the Gaia catalog which
has 46.1\%. The mean SNR improvement
of those stars which do not have negative parallaxes
is 48.6\%.

\subsection{Application Examples}\label{sec:applications}

\begin{figure*}[h]
    \centering
    \includegraphics[width=0.8\textwidth]{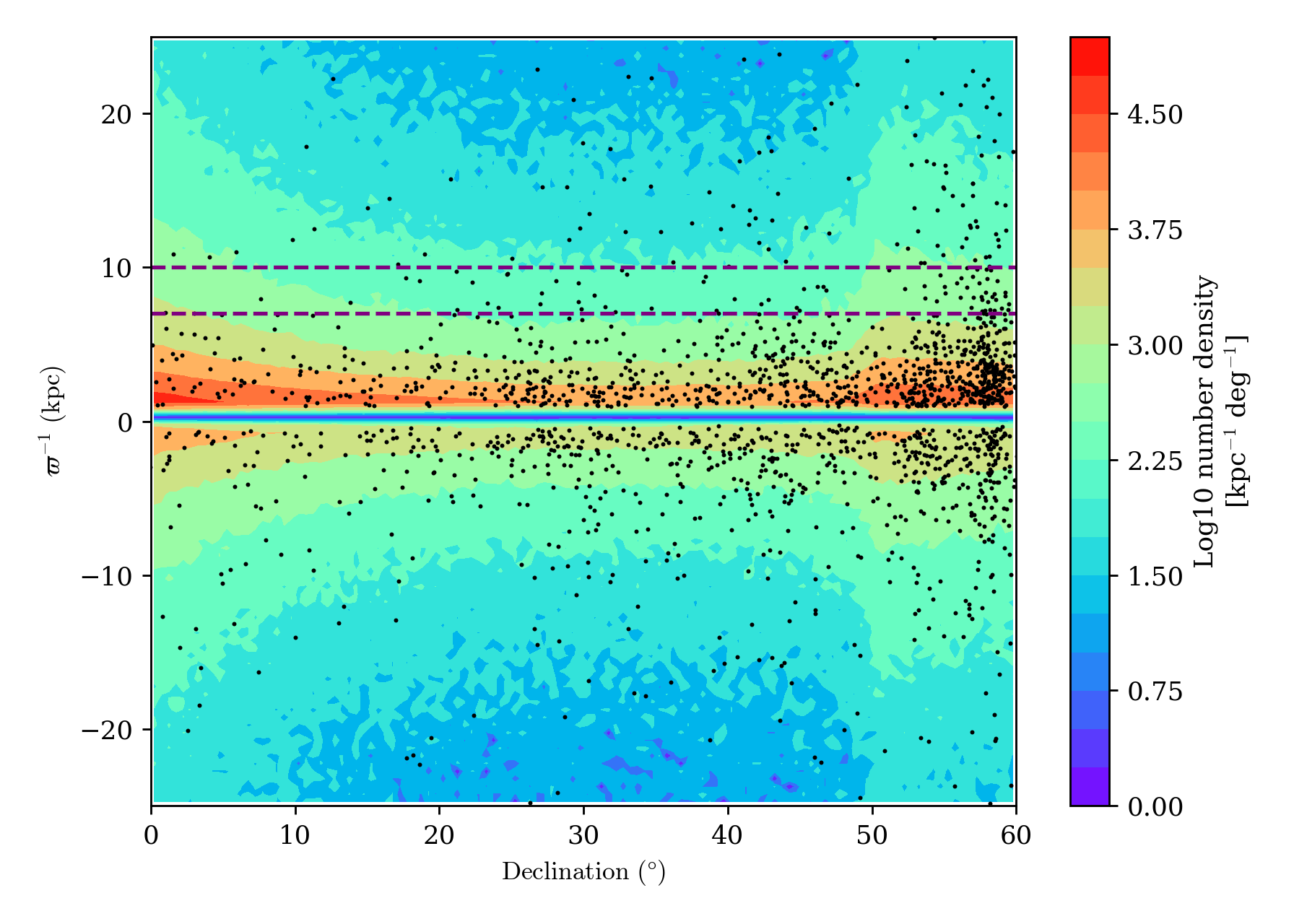}
    \caption{As an example of
    distance estimates without our model,
    the distances to candidate members of the GD-1 stellar stream are shown
    in black dots on the foreground stars as a colored density, shown
    using distance as the inverse of parallax on the y-axis. Compare
    to \cref{fig:gd1_example_prior} and \cref{fig:gd1_example_model}
    as an example of foreground separation without prior knowledge of stellar
    properties. This series of plots demonstrates that the distance prior estimates
    alone are
    too prior dominated to be useful for studying individual clusters of halo stars.}
    \label{fig:gd1_example_dold}
\end{figure*}

\begin{figure*}[h]
    \centering
    \includegraphics[width=0.8\textwidth]{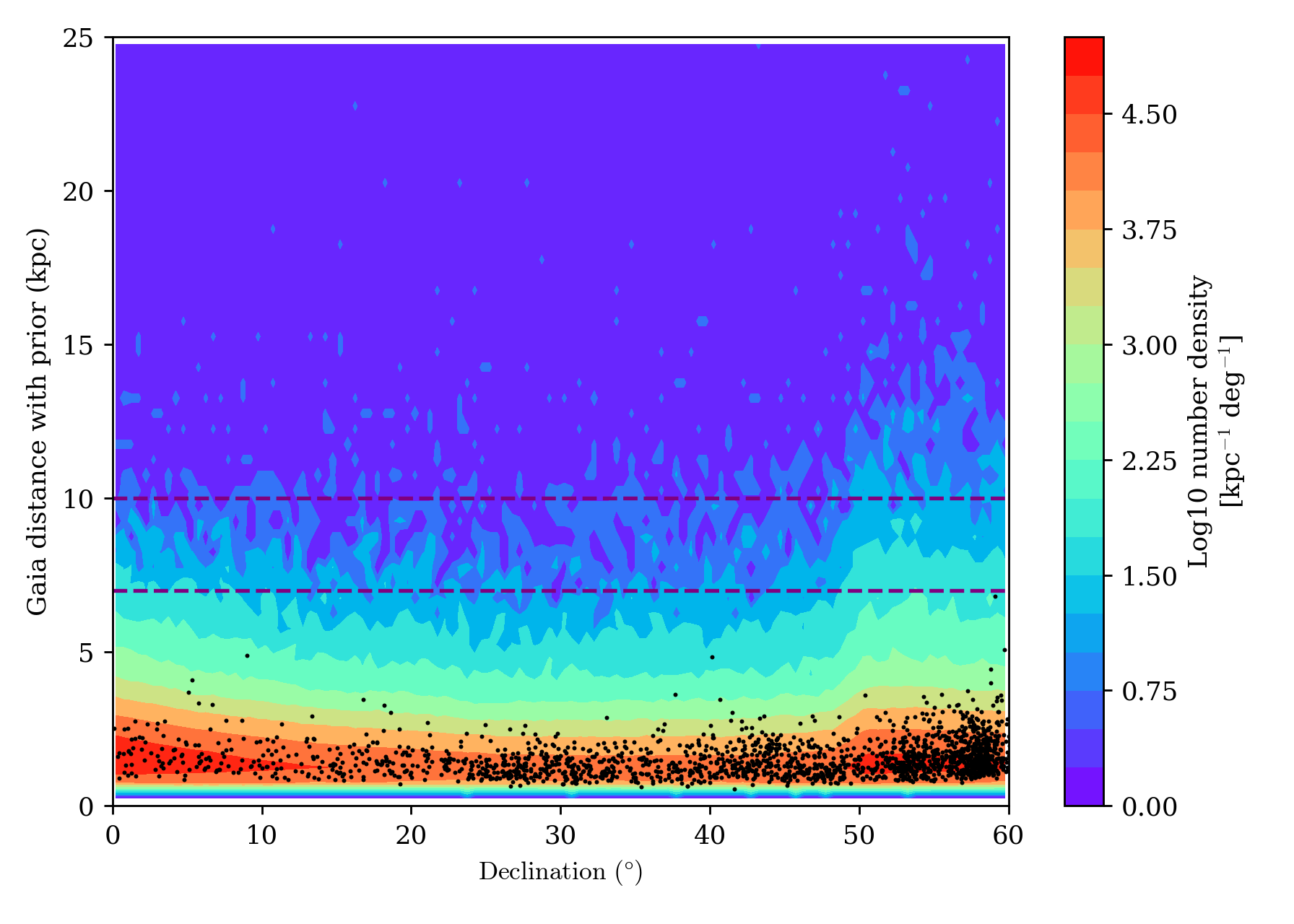}
    \caption{Distances to GD-1 candidate members shown
    in black dots on the foreground stars as a colored density, shown
    using distance priors from \cite{bailer-jones_estimating_2018} with Gaia parallaxes. Compare
    to \cref{fig:gd1_example_dold} and \cref{fig:gd1_example_model}. }
    \label{fig:gd1_example_prior}
\end{figure*}

\begin{figure*}[h]
    \centering
    \includegraphics[width=0.8\textwidth]{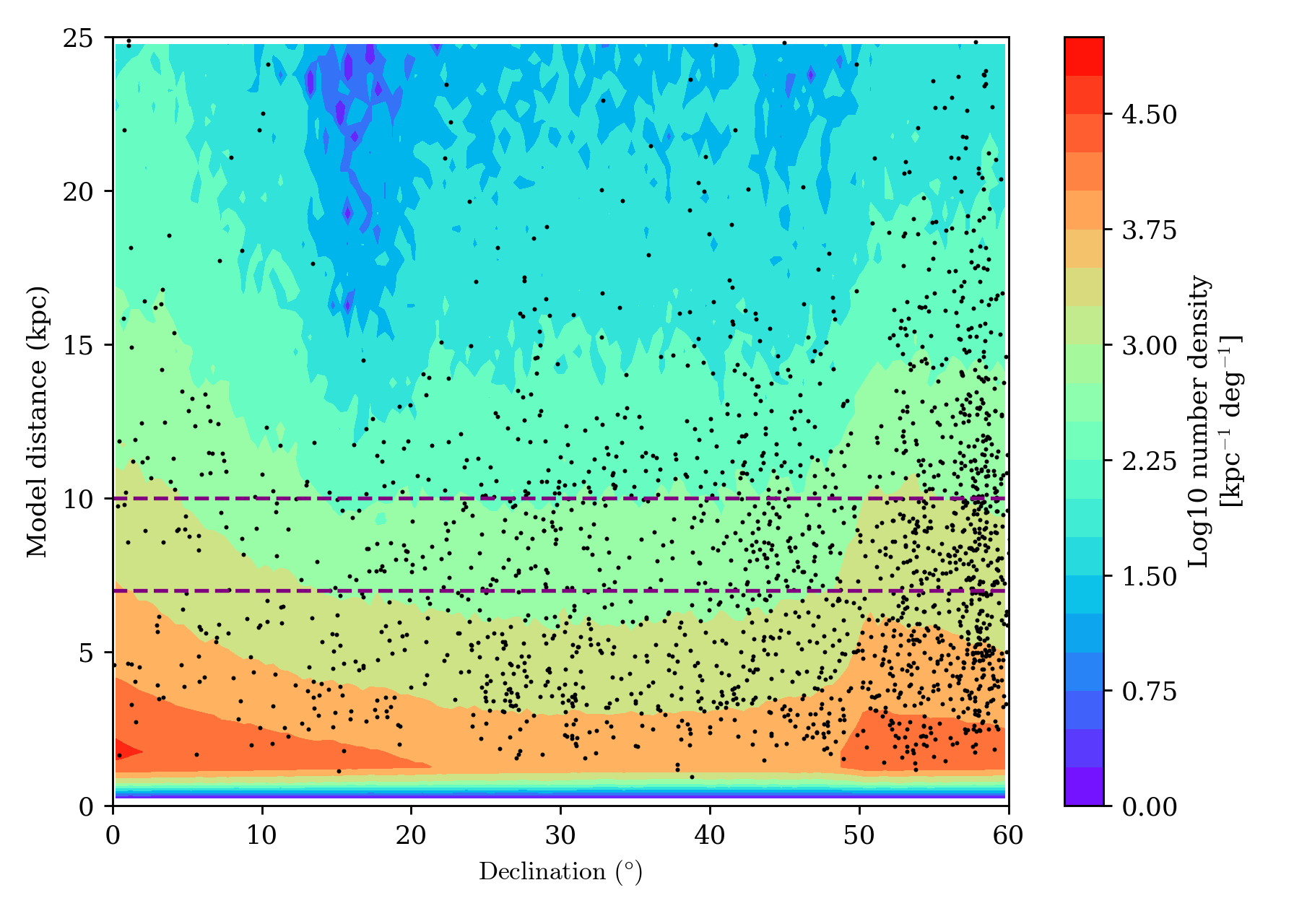}
    \caption{Distances to GD-1 candidate members shown
    in black dots on the foreground stars as a colored density, shown
    using the expectation values of our model. Comparing
    to \cref{fig:gd1_example_dold} and \cref{fig:gd1_example_prior},
    we see that the black dots have moved to bluer contours,
    indicating that the signal-to-noise of the GD-1 stars
    in the declination-distance space improves by orders of magnitude,
    and the typical distance of a candidate star is
    closer to the estimated distance to GD-1 of $\sim$\SI{7}{kpc}.
    With a
    median value of \SI{8}{kpc} and a
    median uncertainty of \SI{6}{kpc},
    these estimates are also more accurate to the
    true distance of GD-1 at about $\sim$7-10 kpc.
    }
    \label{fig:gd1_example_model}
\end{figure*}

Here we describe several potential applications of this
dataset: obtaining an estimate of the distance to M67,
filtering foreground stars, and substructure detection in three spatial
dimensions.
First, we visualize the improved distance estimates to stars
in the GD-1 stream. We make use of the reduction from \cite{price-whelan_off_2018}
to select the stars in a strip of the DR2 sky, and then
compare the distance estimates from Gaia parallaxes alone
with distance estimates from Gaia parallaxes sampled using the Bailer-Jones
distance prior, and finally our model distance. As
seen in \cref{fig:gd1_example_model} compared to \cref{fig:gd1_example_dold} (raw Gaia)
and \cref{fig:gd1_example_prior} (distance priors), distance estimates to stars in
GD-1 are greatly improved with this photometric model.
By using the distance estimates from our model,
one can perform kinematic searches for substructure farther
than $\SI{1}{kpc}$
without requiring associated stars be part of
an isochrone with the same metallicity and age.
In other words, this enables generic filtering of foreground
stars in Gaia data, and also adds a third distance
dimension for clustering stellar substructures, such as with the \texttt{STREAMFINDER} algorithm in \cite{malhan_streamfinder_2018}.

\begin{figure*}[h]
    \centering
    \includegraphics[width=\textwidth]{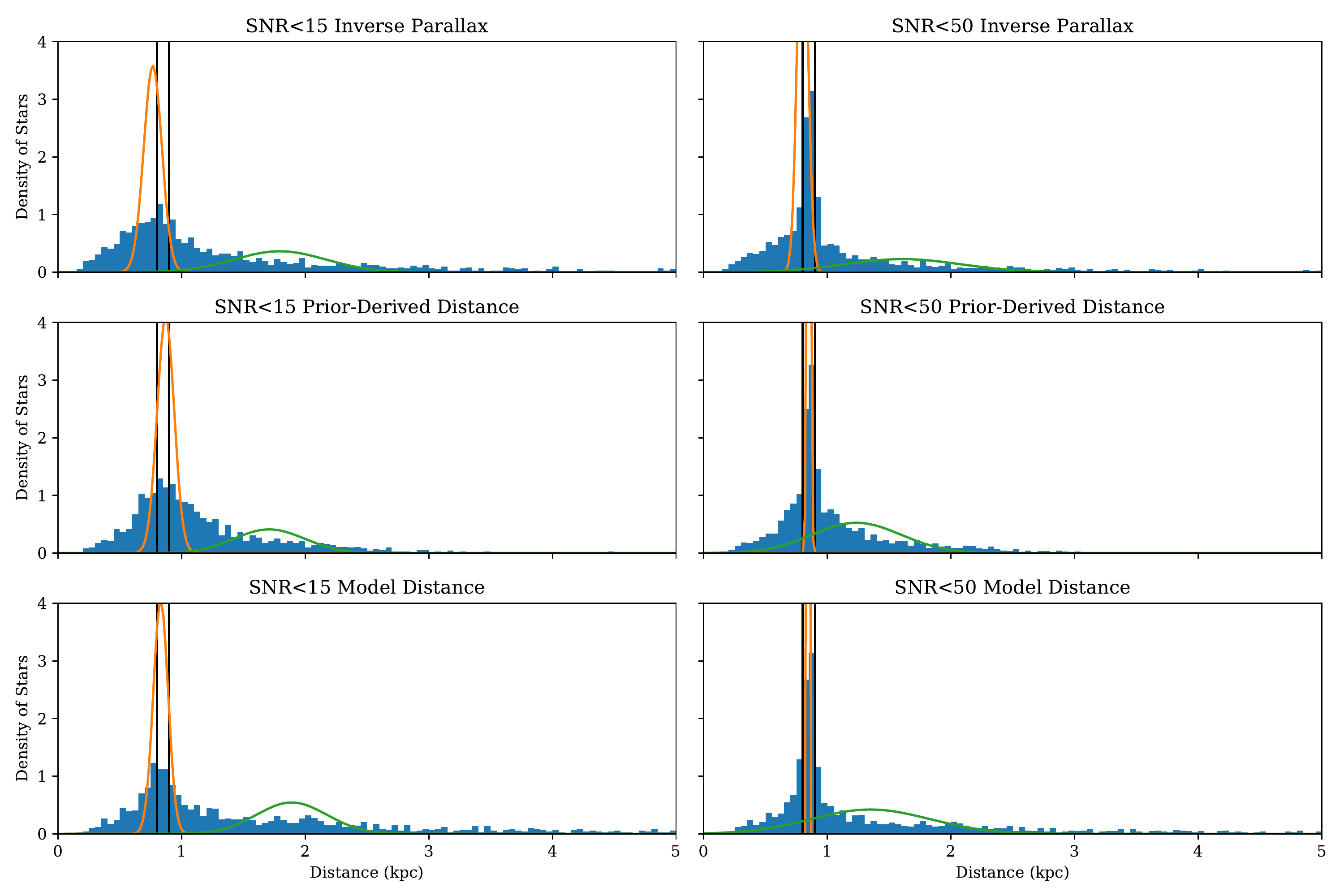}
    \caption{A histogram of Gaia-derived distance estimates to stars within
    \SI{0.3}{\degree} of the direction to the M67 cluster,
    showing only those stars with noisy parallaxes under
    some SNR threshold. The known distance to M67 is within the black lines between 800 and 900 \si{pc}.
    The orange and green Gaussians overplotted show a two-component Bayesian GMM fitted to model
    the M67 cluster and background stars, respectively. Our model
    predicts the distances shown in the bottom row of plots, which are the most
    accurate and precise, as summarized in Table~\ref{tbl:M67}.}
    \label{fig:M67}
\end{figure*}

Finally, we give an example of estimating the distance
to the cluster M67, using full Bayesian posteriors
from our model.
The cluster M67 is at $\sim$\SI{840}{pc} (\citealt{yakut_close_2009})
from Earth
and was also used as a demonstration cluster in \cite{anderson_improving_2018}.
We do this with our model without distance priors (a flat prior
over non-negative distances).
First, we filter Gaia DR2 sources to the cone centered about
$\alpha = 08^\text{h} 51.3^\text{m}$, $\delta$=\ang{11;49;}, with radius \SI{0.3}{\degree}.
Since the estimate of inverted Gaia parallaxes is reasonable by itself (excluding
negative parallaxes),
we filter down to only parallaxes with SNR less than some bound,
shown as the columns in \cref{fig:M67}.
to demonstrate that our model improves the precision and accuracy
of noisy parallaxes.
The top row of \cref{fig:M67} 
shows the inverse parallaxes.
The middle row in \cref{fig:M67} has plots
for the distribution of distances from Gaia parallaxes sampled
using the \cite{bailer-jones_estimating_2018} distance prior,
and then in the bottom row of \cref{fig:M67}, we show the distribution
of the model-derived distances presented in this paper.

We then fit a two-component Bayesian GMM to every distance distribution,
modeling the cluster distance with one component and the background
stars with the other. These are overplotted in \cref{fig:M67}.
The model estimates for each SNR are shown in \cref{tbl:M67}.
\begin{table*}[h]
\centering
\begin{tabular}{@{}lcc@{}}
\toprule
    Model & M67 Center Estimate (kpc) & Spread (kpc) \\\midrule\midrule
    Inverse Parallax with SNR $< 15$ (and $\parobs>0$) & $0.77$ & $0.07$ \\
    Inverse Parallax with SNR $ < 50$ (and $\parobs>0$) & $0.80$ & $0.04$ \\
    Bailer-Jones Prior with SNR $< 15$ & $0.87$ & $0.07$ \\
    Bailer-Jones Prior with SNR $< 50$ & $0.85$ & $0.01$ \\
    Normalizing Flow with SNR $< 15$ & $0.83$ & $0.06$ \\
    Normalizing Flow with SNR $< 50$ & $0.845$ & $0.006$ \\
\bottomrule
\end{tabular}
\caption{This plot
    shows estimates of the distance to the M67 cluster using
    various models on noisy parallax data, for different noise cutoffs.
    SNR is defined here with parallax: $\parobs/\sigma_\varpi$.}
\label{tbl:M67}
\end{table*}
As is evident in \cref{tbl:M67}, our normalizing flow model
improves precision and accuracy
to the M67 cluster of the very low-quality parallaxes:
not only is the spread in the Gaussian component
almost half of that with the
Bailer-Jones distance priors, but the estimate
derived from SNR $<15$ parallaxes is much closer
to the final estimate of $0.84$ kpc using our model.
Also, since the distance priors rely on other surveys to estimate
stellar density in every direction, they implicitly
contain models for this particular cluster, meaning
it is easier for the distance prior model in this example.
The distance prior 
strategy fails in the example in \cref{fig:gd1_example_prior},
with distance estimates to stars in the GD-1 stellar
stream: the estimates are completely off the accepted value range.
When there are anomalous
stars, the prior-derived distance estimates become much less useful,
and a photometry model is necessary, as shown
giving much more reasonable distances in \cref{fig:gd1_example_model}.

\section{Future work}\label{sec:future}

The Masked Autoencoder architecture from \cite{germain_made:_2015}
and \cite{papamakarios_masked_2017} which
we use in our technique models a joint posterior in $\mathbb{R}^n$ using conditional probabilities:
\begin{align*}
    p(x_1, \ldots, x_n) &\sim p(x_1) p(x_2|x_1) p(x_3| x_1, x_2)\cdots p(x_n| x_1, \ldots x_{n-1}).
\end{align*}
What this means is that we can exactly marginalize over $x_{n-m}$ through 
$x_n$ for some $m$ without additional computation by excluding
them from the joint posterior, as long as
we fix a hierarchy of $x_i$ that maximizes.
In other words, the marginalized probability is:
\begin{align*}
    p(x_1, \ldots, x_{n-m}) &\sim p(x_1) p(x_2|x_1) p(x_3| x_1, x_2) \cdots p(x_{n-m}| x_1, \ldots x_{n-m-1}),
\end{align*}
which uses the same normalizing flow model.
While we cannot marginalize over an arbitrary $x_i$ without using an ensemble model, 
if we have photometric bands with different coverage, we can fit a 
color-magnitude density that simultaneously models many different
survey bands, and still train and evaluate the color-magnitude probability
with limited information. E.g., if we were to model Gaia and AllWISE
bands simultaneously in a color-magnitude density, and we order Gaia $>$ AllWISE,
then we can exactly evaluate the color-magnitude density for a given Gaia photometry
marginalized over AllWISE bands, but not vice versa.
This technique, also used in \cite{alsing_fast_2019}, is
powerful and would be useful to exploit for future distance estimates
to maximize coverage while using all available surveys.
In the future, we would also like to make use of maps for $R_V$, 
as well as simultaneously model stellar parameters, rather than use
extinction conversions based on a simple blackbody.

\section{Conclusion}
\label{sec:discussion}

We have demonstrated an algorithm for learning a flexible
probability distribution in Gaia
color-magnitude space from noisy parallax
and photometry measurements
using a normalizing flow.
These deep
neural networks, capable of learning
arbitrary multi-dimensional probability distributions,
have been shown in this paper to be capable of
modeling CMDs well, and work at predicting CMDs accurately
in an iterative
dust estimation scheme.
We have also presented a catalog of 640M photometric distance
posteriors derived from this data-driven model
using Gaia DR2 photometry and parallaxes to learn
a prior in Gaia color space.
Overall, the signal-to-noise
of distance measurements in this catalog improves on average by
48\% over the raw Gaia data, including only 
the non-negative Gaia parallaxes, and we also
demonstrate how the accuracy of distances have improved over
other models.
Applications are discussed for this catalog, including significantly improved Milky Way
disk separation and substructure detection.

We also will maintain
a GitHub repository at \url{https://github.com/MilesCranmer/public_CMD_normalizing_flow}
where we will post links to the distance catalog along with future versions,
host code of the normalizing flow for photometry data, and
answer questions about the paper and implementing the algorithm.

\section{Acknowledgments}

Miles Cranmer would like to thank David W. Hogg
for the initial project idea of improving
Gaia distances with a generative neural network model,
Johann Brehmer and Thomas Kipf
for the idea of creating this model
with a normalizing flow,
Wolfgang Kerzendorf
for feedback on calculating extinction conversions,
Gregory Green for help with his \texttt{dustmap} package,
George Papamakarios for the idea of marginalizing
over input to the normalizing flow using
\cite{papamakarios_masked_2017},
and Iain Murray for comments on an early draft.
This work made use of Astropy (\citealt{robitaille_astropy:_2013}),
PyTorch (\citealt{paszke2017automatic}), Scikit-Learn (\citealt{scikit-learn}),
among other scientific packages mentioned in the paper.

\bibliography{main}

\begin{thebibliography}{}
\expandafter\ifx\csname natexlab\endcsname\relax\def\natexlab#1{#1}\fi
\providecommand{\url}[1]{\href{#1}{#1}}
\providecommand{\dodoi}[1]{doi:~\href{http://doi.org/#1}{\nolinkurl{#1}}}
\providecommand{\doeprint}[1]{\href{http://ascl.net/#1}{\nolinkurl{http://ascl.net/#1}}}
\providecommand{\doarXiv}[1]{\href{https://arxiv.org/abs/#1}{\nolinkurl{https://arxiv.org/abs/#1}}}

\bibitem[{Alsing {et~al.}(2019)Alsing, Charnock, Feeney, \&
  Wandelt}]{alsing_fast_2019}
Alsing, J., Charnock, T., Feeney, S., \& Wandelt, B. 2019, \mnras, stz1960,
  \dodoi{10.1093/mnras/stz1960}

\bibitem[{Anderson {et~al.}(2018)Anderson, Hogg, Leistedt, Price-Whelan, \&
  Bovy}]{anderson_improving_2018}
Anderson, L., Hogg, D.~W., Leistedt, B., Price-Whelan, A.~M., \& Bovy, J. 2018,
  \aj, 156, 145, \dodoi{10.3847/1538-3881/aad7bf}

\bibitem[{Bailer-Jones(2015)}]{bailer-jones_estimating_2015}
Bailer-Jones, C. A.~L. 2015, \pasp, 127, 994, \dodoi{10.1086/683116}

\bibitem[{Bailer-Jones {et~al.}(2018)Bailer-Jones, Rybizki, Fouesneau,
  Mantelet, \& Andrae}]{bailer-jones_estimating_2018}
Bailer-Jones, C. a.~L., Rybizki, J., Fouesneau, M., Mantelet, G., \& Andrae, R.
  2018, \aap, 156, 58, \dodoi{10.3847/1538-3881/aacb21}

\bibitem[{Barbary(2016)}]{barbary_extinction_2016}
Barbary, K. 2016, extinction v0.3.0, \dodoi{10.5281/zenodo.804967}.
\newblock \url{https://doi.org/10.5281/zenodo.804967}

\bibitem[{Bonaca {et~al.}(2019)Bonaca, Hogg, Price-Whelan, \&
  Conroy}]{bonaca_spur_2019}
Bonaca, A., Hogg, D.~W., Price-Whelan, A.~M., \& Conroy, C. 2019, \apj, 880,
  38, \dodoi{10.3847/1538-4357/ab2873}

\bibitem[{Bovy(2017)}]{bovy_stellar_2017}
Bovy, J. 2017, \mnras, 470, 1360, \dodoi{10.1093/mnras/stx1277}

\bibitem[{{Bovy Jo} {et~al.}(2011){Bovy Jo}, Hogg, \&
  Roweis}]{bovy_jo_extreme_2011}
{Bovy Jo}, Hogg, D.~W., \& Roweis, S.~T. 2011, Annals of Applied Statistics, 5,
  1657, \dodoi{10.1214/10-AOAS439}

\bibitem[{Brehmer {et~al.}(2019)Brehmer, Kling, Espejo, \&
  Cranmer}]{brehmer_madminer:_2019}
Brehmer, J., Kling, F., Espejo, I., \& Cranmer, K. 2019, arXiv e-prints,
  arXiv:1907.10621

\bibitem[{Brown {et~al.}(2018)Brown, Vallenari, Prusti, Bruijne, Babusiaux,
  Bailer-Jones, Biermann, Evans, Eyer, Jansen, Jordi, Klioner, Lammers,
  Lindegren, Luri, Mignard, Panem, Pourbaix, Randich, Sartoretti, Siddiqui,
  Soubiran, Leeuwen, Walton, Arenou, Bastian, Cropper, Drimmel, Katz, Lattanzi,
  Bakker, Cacciari, Castañeda, Chaoul, Cheek, Angeli, Fabricius, Guerra, Holl,
  Masana, Messineo, Mowlavi, Nienartowicz, Panuzzo, Portell, Riello, Seabroke,
  Tanga, Thévenin, Gracia-Abril, Comoretto, Garcia-Reinaldos, Teyssier,
  Altmann, Andrae, Audard, Bellas-Velidis, Benson, Berthier, Blomme, Burgess,
  Busso, Carry, Cellino, Clementini, Clotet, Creevey, Davidson, Ridder,
  Delchambre, Dell’Oro, Ducourant, Fernández-Hernández, Fouesneau, Frémat,
  Galluccio, García-Torres, González-Núñez, González-Vidal, Gosset, Guy,
  Halbwachs, Hambly, Harrison, Hernández, Hestroffer, Hodgkin, Hutton,
  Jasniewicz, Jean-Antoine-Piccolo, Jordan, Korn, Krone-Martins, Lanzafame,
  Lebzelter, Löffler, Manteiga, Marrese, Martín-Fleitas, Moitinho, Mora,
  Muinonen, Osinde, Pancino, Pauwels, Petit, Recio-Blanco, Richards, Rimoldini,
  Robin, Sarro, Siopis, Smith, Sozzetti, Süveges, Torra, Reeven, Abbas,
  Aramburu, Accart, Aerts, Altavilla, Álvarez, Alvarez, Alves, Anderson,
  Andrei, Varela, Antiche, Antoja, Arcay, Astraatmadja, Bach, Baker,
  Balaguer-Núñez, Balm, Barache, Barata, Barbato, Barblan, Barklem, Barrado,
  Barros, Barstow, Muñoz, Bassilana, Becciani, Bellazzini, Berihuete, Bertone,
  Bianchi, Bienaymé, Blanco-Cuaresma, Boch, Boeche, Bombrun, Borrachero,
  Bossini, Bouquillon, Bourda, Bragaglia, Bramante, Breddels, Bressan,
  Brouillet, Brüsemeister, Brugaletta, Bucciarelli, Burlacu, Busonero,
  Butkevich, Buzzi, Caffau, Cancelliere, Cannizzaro, Cantat-Gaudin, Carballo,
  Carlucci, Carrasco, Casamiquela, Castellani, Castro-Ginard, Charlot, Chemin,
  Chiavassa, Cocozza, Costigan, Cowell, Crifo, Crosta, Crowley, Cuypers†,
  Dafonte, Damerdji, Dapergolas, David, David, Laverny, Luise, March, Martino,
  Souza, Torres, Debosscher, Pozo, Delbo, Delgado, Delgado, Matteo, Diakite,
  Diener, Distefano, Dolding, Drazinos, Durán, Edvardsson, Enke, Eriksson,
  Esquej, Bontemps, Fabre, Fabrizio, Faigler, Falcão, Casas, Federici,
  Fedorets, Fernique, Figueras, Filippi, Findeisen, Fonti, Fraile, Fraser,
  Frézouls, Gai, Galleti, Garabato, García-Sedano, Garofalo, Garralda, Gavel,
  Gavras, Gerssen, Geyer, Giacobbe, Gilmore, Girona, Giuffrida, Glass, Gomes,
  Granvik, Gueguen, Guerrier, Guiraud, Gutiérrez-Sánchez, Haigron,
  Hatzidimitriou, Hauser, Haywood, Heiter, Helmi, Heu, Hilger, Hobbs, Hofmann,
  Holland, Huckle, Hypki, Icardi, Janßen, Fombelle, Jonker, Juhász, Julbe,
  Karampelas, Kewley, Klar, Kochoska, Kohley, Kolenberg, Kontizas, Kontizas,
  Koposov, Kordopatis, Kostrzewa-Rutkowska, Koubsky, Lambert, Lanza, Lasne,
  Lavigne, Fustec, Poncin-Lafitte, Lebreton, Leccia, Leclerc, Lecoeur-Taibi,
  Lenhardt, Leroux, Liao, Licata, Lindstrøm, Lister, Livanou, Lobel, López,
  Managau, Mann, Mantelet, Marchal, Marchant, Marconi, Marinoni, Marschalkó,
  Marshall, Martino, Marton, Mary, Massari, Matijevič, Mazeh, McMillan,
  Messina, Michalik, Millar, Molina, Molinaro, Molnár, Montegriffo, Mor,
  Morbidelli, Morel, Morris, Mulone, Muraveva, Musella, Nelemans, Nicastro,
  Noval, O’Mullane, Ordénovic, Ordóñez-Blanco, Osborne, Pagani, Pagano,
  Pailler, Palacin, Palaversa, Panahi, Pawlak, Piersimoni, Pineau, Plachy,
  Plum, Poggio, Poujoulet, Prša, Pulone, Racero, Ragaini, Rambaux,
  Ramos-Lerate, Regibo, Reylé, Riclet, Ripepi, Riva, Rivard, Rixon, Roegiers,
  Roelens, Romero-Gómez, Rowell, Royer, Ruiz-Dern, Sadowski, Sellés,
  Sahlmann, Salgado, Salguero, Sanna, Santana-Ros, Sarasso, Savietto,
  Schultheis, Sciacca, Segol, Segovia, Ségransan, Shih, Siltala, Silva, Smart,
  Smith, Solano, Solitro, Sordo, Nieto, Souchay, Spagna, Spoto, Stampa, Steele,
  Steidelmüller, Stephenson, Stoev, Suess, Surdej, Szabados, Szegedi-Elek,
  Tapiador, Taris, Tauran, Taylor, Teixeira, Terrett, Teyssandier, Thuillot,
  Titarenko, Clotet, Turon, Ulla, Utrilla, Uzzi, Vaillant, Valentini, Valette,
  Elteren, Hemelryck, Leeuwen, Vaschetto, Vecchiato, Veljanoski, Viala,
  Vicente, Vogt, Essen, Voss, Votruba, Voutsinas, Walmsley, Weiler, Wertz,
  Wevers, Wyrzykowski, Yoldas, Žerjal, Ziaeepour, Zorec, Zschocke, Zucker,
  Zurbach, \& Zwitter}]{brown_gaia_2018}
Brown, A. G.~A., Vallenari, A., Prusti, T., {et~al.} 2018, \aap, 616, A1,
  \dodoi{10.1051/0004-6361/201833051}

\bibitem[{Dinh {et~al.}(2016)Dinh, Sohl-Dickstein, \&
  Bengio}]{dinh_density_2016}
Dinh, L., Sohl-Dickstein, J., \& Bengio, S. 2016, arXiv:1605.08803 [cs, stat]

\bibitem[{Germain {et~al.}(2015)Germain, Gregor, Murray, \&
  Larochelle}]{germain_made:_2015}
Germain, M., Gregor, K., Murray, I., \& Larochelle, H. 2015, arXiv:1502.03509
  [cs, stat]

\bibitem[{Glorot \& Bengio(2010)}]{glorot_understanding_2010}
Glorot, X., \& Bengio, Y. 2010, in Proceedings of the thirteenth international
  conference on artificial intelligence and statistics, 249--256

\bibitem[{{Green}(2018)}]{dustmaps}
{Green}, G. 2018, The Journal of Open Source Software, 3, 695,
  \dodoi{10.21105/joss.00695}

\bibitem[{Green {et~al.}(2019)Green, Schlafly, Zucker, Speagle, \&
  Finkbeiner}]{green_3d_2019}
Green, G.~M., Schlafly, E.~F., Zucker, C., Speagle, J.~S., \& Finkbeiner, D.~P.
  2019, arXiv e-prints, arXiv:1905.02734

\bibitem[{Green {et~al.}(2018)Green, Schlafly, Finkbeiner, Rix, Martin,
  Burgett, Draper, Flewelling, Hodapp, Kaiser, Kudritzki, Magnier, Metcalfe,
  Tonry, Wainscoat, \& Waters}]{green_galactic_2018}
Green, G.~M., Schlafly, E.~F., Finkbeiner, D., {et~al.} 2018, \mnras, 478, 651,
  \dodoi{10.1093/mnras/sty1008}

\bibitem[{Hall {et~al.}(2019)Hall, Davies, Elsworth, Miglio, Bedding, Brown,
  Khan, Hawkins, García, Chaplin, \& North}]{hall_testing_2019}
Hall, O.~J., Davies, G.~R., Elsworth, Y.~P., {et~al.} 2019, \mnras, 486, 3569,
  \dodoi{10.1093/mnras/stz1092}

\bibitem[{Hawkins {et~al.}(2017)Hawkins, Leistedt, Bovy, \&
  Hogg}]{hawkins_red_2017}
Hawkins, K., Leistedt, B., Bovy, J., \& Hogg, D.~W. 2017, \mnras, 471, 722,
  \dodoi{10.1093/mnras/stx1655}

\bibitem[{Hogg(2018)}]{hogg_likelihood_2018}
Hogg, D.~W. 2018, ArXiv e-prints, 1804, arXiv:1804.07766

\bibitem[{Huber {et~al.}(2017)Huber, Zinn, Bojsen-Hansen, Pinsonneault,
  Sahlholdt, Serenelli, Silva~Aguirre, Stassun, Stello, Tayar, Bastien,
  Bedding, Buchhave, Chaplin, Davies, García, Latham, Mathur, Mosser, \&
  Sharma}]{huber_asteroseismology_2017}
Huber, D., Zinn, J., Bojsen-Hansen, M., {et~al.} 2017, \apj, 844, 102,
  \dodoi{10.3847/1538-4357/aa75ca}

\bibitem[{Koposov {et~al.}(2019)Koposov, Belokurov, Li, Mateu, Erkal,
  Grillmair, Hendel, Price-Whelan, Laporte, Hawkins, Sohn, del Pino, Evans,
  Slater, Kallivayalil, Navarro, \& {(The OATs: Orphan Aspen Treasury
  Collaboration)}}]{koposov_piercing_2019}
Koposov, S.~E., Belokurov, V., Li, T.~S., {et~al.} 2019, \mnras, 485, 4726,
  \dodoi{10.1093/mnras/stz457}

\bibitem[{Leistedt \& Hogg(2017)}]{leistedt_hierarchical_2017}
Leistedt, B., \& Hogg, D.~W. 2017, \aap, 154, 222,
  \dodoi{10.3847/1538-3881/aa91d5}

\bibitem[{Lindegren {et~al.}(2018)Lindegren, Hernández, Bombrun, Klioner,
  Bastian, Ramos-Lerate, Torres, Steidelmüller, Stephenson, Hobbs, Lammers,
  Biermann, Geyer, Hilger, Michalik, Stampa, McMillan, Castañeda, Clotet,
  Comoretto, Davidson, Fabricius, Gracia, Hambly, Hutton, Mora, Portell,
  Leeuwen, Abbas, Abreu, Altmann, Andrei, Anglada, Balaguer-Núñez, Barache,
  Becciani, Bertone, Bianchi, Bouquillon, Bourda, Brüsemeister, Bucciarelli,
  Busonero, Buzzi, Cancelliere, Carlucci, Charlot, Cheek, Crosta, Crowley,
  Bruijne, Felice, Drimmel, Esquej, Fienga, Fraile, Gai, Garralda,
  González-Vidal, Guerra, Hauser, Hofmann, Holl, Jordan, Lattanzi, Lenhardt,
  Liao, Licata, Lister, Löffler, Marchant, Martin-Fleitas, Messineo, Mignard,
  Morbidelli, Poggio, Riva, Rowell, Salguero, Sarasso, Sciacca, Siddiqui,
  Smart, Spagna, Steele, Taris, Torra, Elteren, Reeven, \&
  Vecchiato}]{lindegren_gaia_2018}
Lindegren, L., Hernández, J., Bombrun, A., {et~al.} 2018, \aap, 616, A2,
  \dodoi{10.1051/0004-6361/201832727}

\bibitem[{Malhan \& Ibata(2018)}]{malhan_streamfinder_2018}
Malhan, K., \& Ibata, R.~A. 2018, \mnras, 477, 4063,
  \dodoi{10.1093/mnras/sty912}

\bibitem[{Malhan {et~al.}(2018)Malhan, Ibata, \& Martin}]{malhan_ghostly_2018}
Malhan, K., Ibata, R.~A., \& Martin, N.~F. 2018, \mnras, 481, 3442,
  \dodoi{10.1093/mnras/sty2474}

\bibitem[{O'Donnell(1994)}]{odonnell_rnu-dependent_1994}
O'Donnell, J.~E. 1994, \apj, 422, 158, \dodoi{10.1086/173713}

\bibitem[{Papamakarios {et~al.}(2017)Papamakarios, Pavlakou, \&
  Murray}]{papamakarios_masked_2017}
Papamakarios, G., Pavlakou, T., \& Murray, I. 2017, arXiv e-prints,
  arXiv:1705.07057

\bibitem[{Papamakarios {et~al.}(2018)Papamakarios, Sterratt, \&
  Murray}]{papamakarios_sequential_2018}
Papamakarios, G., Sterratt, D.~C., \& Murray, I. 2018, arXiv:1805.07226 [cs,
  stat]

\bibitem[{Paszke {et~al.}(2017)Paszke, Gross, Chintala, Chanan, Yang, DeVito,
  Lin, Desmaison, Antiga, \& Lerer}]{paszke2017automatic}
Paszke, A., Gross, S., Chintala, S., {et~al.} 2017, in NIPS Autodiff Workshop

\bibitem[{Pedregosa {et~al.}(2011)Pedregosa, Varoquaux, Gramfort, Michel,
  Thirion, Grisel, Blondel, Prettenhofer, Weiss, Dubourg, Vanderplas, Passos,
  Cournapeau, Brucher, Perrot, \& Duchesnay}]{scikit-learn}
Pedregosa, F., Varoquaux, G., Gramfort, A., {et~al.} 2011, Journal of Machine
  Learning Research, 12, 2825

\bibitem[{Price-Whelan \& Bonaca(2018)}]{price-whelan_off_2018}
Price-Whelan, A.~M., \& Bonaca, A. 2018, \apj, 863, L20,
  \dodoi{10.3847/2041-8213/aad7b5}

\bibitem[{Riello {et~al.}(2018)Riello, Angeli, Evans, Busso, Hambly, Davidson,
  Burgess, Montegriffo, Osborne, Kewley, Carrasco, Fabricius, Jordi, Cacciari,
  Leeuwen, \& Holland}]{riello_gaia_2018}
Riello, M., Angeli, F.~D., Evans, D.~W., {et~al.} 2018, \aap, 616, A3,
  \dodoi{10.1051/0004-6361/201832712}

\bibitem[{Robitaille {et~al.}(2013)Robitaille, Tollerud, Greenfield,
  Droettboom, Bray, Aldcroft, Davis, Ginsburg, Price-Whelan, Kerzendorf,
  Conley, Crighton, Barbary, Muna, Ferguson, Grollier, Parikh, Nair, Günther,
  Deil, Woillez, Conseil, Kramer, Turner, Singer, Fox, Weaver, Zabalza,
  Edwards, Bostroem, Burke, Casey, Crawford, Dencheva, Ely, Jenness, Labrie,
  Lim, Pierfederici, Pontzen, Ptak, Refsdal, Servillat, \&
  Streicher}]{robitaille_astropy:_2013}
Robitaille, T.~P., Tollerud, E.~J., Greenfield, P., {et~al.} 2013, \aap, 558,
  A33, \dodoi{10.1051/0004-6361/201322068}

\bibitem[{Yakut {et~al.}(2009)Yakut, Zima, Kalomeni, van Winckel, Waelkens,
  De~Cat, Bauwens, Vučković, Saesen, Le~Guillou, Parmaksızoğlu, Uluç,
  Khamitov, Raskin, \& Aerts}]{yakut_close_2009}
Yakut, K., Zima, W., Kalomeni, B., {et~al.} 2009, \aap, 503, 165,
  \dodoi{10.1051/0004-6361/200911918}

\end{thebibliography}

\end{document}